\documentclass{article}
\usepackage{sao2}
\usepackage{psfig}
\issue{2001, 52, 5-133}
\begin{document}
\markboth{Verkhodanov, Parijskij, Soboleva, Kopylov, Temirova, Zhelenkova,
Goss}{Results of investigation of radiogalaxies of the survey ``Cold''}
\title{Results of investigation of radio galaxies of the survey ``Cold'':
photometry, colour redshifts and the age of the stellar population}
\author{O.V. Verkhodanov\inst{a} \and Yu.N. Parijskij\inst{a} \and
N.S. Soboleva\inst{b} \and
A.I. Kopylov\inst{a} \and A.V. Temirova\inst{b} \and O.P. Zhelenkova\inst{a}
 \and W.M. Goss\inst{c}}
\institute{\saoname \and St.Petersburg branch of SAO RAS, St.Petersburg, Russia
\and National Radio Astronomical Observatory, Soccoro, USA}
\date{November 14, 2001}{December 1, 2001}
\maketitle
\begin{abstract}
BVRI data for the majority of the objects of the RC catalogue with steep
spectra and $m_R<24^m$ are presented. These data have been used to estimate
colour redshifts and the age of stellar systems of host galaxies. By way of
example of distant radio galaxies it is shown that this approach gives an
accuracy of redshift estimates close to that for field galaxies
($\sim20$\,\%). The age estimates are less confident, however, the lower
age limit for not too distant ($z<1.5$) objects is determined quite reliably.
Several galaxies have been detected that have an age above that of the
Universe at the given $z$ in a simple CDM model of the Universe. A possibility
of using such objects to specify the part played by ``dark energy'' is
discussed. This paradox disappears in a model with the $\Lambda$-term equal
to 0.6--0.8.
\end{abstract}

\section{Introduction}
A population of powerful radio galaxies is observable with the
available facilities practically to any distances. This allows one to follow
the evolution of this population in the radio range from the moment it
originated up to the present time.

However, a good deal of effort has to be undertaken to find new objects at
$z>3$ with application of rigorous methods of selection by radio, optical and
infrared properties, which leads to inhomogeneous samples with regard to
different parameters.

It is customary to assume that a population of very powerful radio galaxies
of type FR\,II owes its origin to giant elliptical galaxies having
supermassive ($10^9M_{\sun}$) black holes at the centre. That is why the
evolution of this population is related to the problem of formation
of the largest stellar systems and to the problem of evolution of massive
black holes. Besides, these objects are often associated with clusters and
groups of galaxies and can be indicators of distant clusters formed in the
nodes of the large-scale structure. At last, there is a number of suggestions
for using them in estimation of parameters of their environment and even of
the geometry of the Universe and its dynamics (Parijskij et al., 1998; Daly,
1994).

In contrast to quasar, in radio galaxies one can study in details the stellar population. As a
rule, in objects with medium and weak energy release of the active nucleus
in the optical and near infrared ranges, radiation of stars rather than gas
dominates at $z<1-1.5$. For this reason, one may attempt to apply the method
of stellar evolution and evolution of synthetic colours of the stellar
population to determination of ``colour'' redshifts and the age of the stellar
population.
When lucky, one can define the moment  of
star formation ($z$).

In the generally accepted scheme of the Unified Theory of
active objects, radio galaxies of type FR\,II differ from quasars, BL\,Lac
objects and other populations with powerful nucleus activity only in that
at what angle the axis of rotation
of the gas-dust tore, that screens the accretion disk around the black hole,
is located with respect to the observer. And the data on  evolution of
these objects can be applied to other classes of objects.

At last, powerful galaxies with active nuclei (radio galaxies and quasars,
predominantly ``radio quiet'') supply much UV radiation to the Universe,
and in combination with the integral optical depth by Thomson scattering
of relic photons of the 3K background permit the moment of secondary ionization
of the Universe to be refined.

Deep optical investigations of host galaxies in this class of objects are
impeded by the fact that their spatial density is by 5--6 orders of magnitude
less than that of background galaxies, and this is why they are virtually
lacking in ultimately deep frames of small size. In the HDF (Hubble Deep
Fields) there is found but one distant ($z=4.42$) radio galaxy of medium power.
For this reason, sampling of fields for such investigations has to be
started with preliminary selection on radio astronomy data with allowance
made for all possible indirect criteria. The principles of selection of
candidates for distant objects in the ``Big Trio'' project (Parijskij et al.,
1994, 1096a,b) are similar to universally adopted (McCarthy, 1993).
The list of objects presented below has been derived from the catalogue of the
``Cold'' survey (Parijskij et al., 1991, 1992) as a result of several
selection steps: selection of comparatively faint sources having steep
spectra ($\alpha<-1.0 S\sim\nu^{\alpha}$) and being of the FR\,II
morphological type (Fanaroff, Riley, 1974)
 in the cm range (10--100\,mJy). Further selection was done according to the
procedure described below.

The labour
input for acquisition of quality spectroscopic data on distant and faint
galaxies and radio galaxies  forces to search for indirect methods for determination
of the redshift and other characteristic features of these objects. With regard
to the powerful  radio galaxies, even photometric estimates proved to be
helpful and they  have so far been widely used (McCarthy, 1993; Benn et al.,
1989; Parijskij et al., 2000a,\,b).

Over the last few years, colour characteristics of faint galaxies have come
to be used in addition to the methods of photometric evaluation of the
redshift, and this approach forms a basis for a number of major projects
(e.g. see Szalay, 1996). As we have already noted, the colours of stellar
systems make it possible to estimate their age as well. Generally speaking,
the legitimacy of using colours for such estimations of evolution
characteristics of the population of powerful radio galaxies requires a
separate study because the effect the nuclear activity in these objects has
on the colour characteristics is not sufficiently understood. The sites of
``secondary'' star formation may also tell upon the colour characteristics.

The present paper is a continuation of our programme of investigation of
the radio galaxies detected in the RATAN--600 survey ``Cold'' with involvement
of multicolour photometry data for assessment of colour redshifts and ages of
stellar systems of host galaxies. Current use of these data is practically
inevitable. Direct spectroscopy of faint objects required until recently
a great deal of the 6\,m telescope observational time when working with
objects fainter than $20^m$. So, in 1995--1996 we managed to measure the
spectroscopic redshift but in four bright objects (three quasars and one
galaxy, all being brighter than $m_R=20.5^m$) and only in one faint galaxy,
$m_R=23^m$ (Dodonov et al., 1999). The redshift of the latter, $z=2.73$,
was estimated from the only emission line. It is independently confirmed
by our colour data: the negative B--V colour index agrees with identification
of a strong line as $Ly_{\alpha}$, which is typical of radio galaxies.

During the last time, in connection with placing in service of more efficient
spectral tools in SAO RAS, spectroscopy of faint objects is expected to
substantionally improve. The spectra of another two objects, radio galaxy
RCJ\,0908+0451 and quasar RCJ\,1154+0431, were obtained at the 6\,m telescope
in 2001 March with the aid of new equipment SCORPIO (Afanasiev and Moiseev,
{\it http://www.sao.ru/$\sim$moisav/scorpio/scorpio.html}). The spectral
redshift for the radio galaxy is practically the same as the colour redshift.
The data on the colour redshift will help improve the efficiency of spectroscopic
observations.

This paper has the following structure. In the first section we present the
data of multicolour BVRI photometry and also the latest results of
identifications of the ``Cold'' survey radio sources. Further we describe
the procedures of obtaining colour redshifts and ages of stellar systems
from BVRI magnitudes with the use of present-day models of spectral energy
distribution (SED). In the sections that follow we discuss the results
obtained, evaluate model differences in the estimates of redshifts and
propose new steps to improve the accuracy of estimations.

\section{Multicolour photometry of RC objects}
By the present time observations of about 60 RC objects (radio galaxies and
quasars) of our sample have been carried out with the 6\,m telescope of
SAO RAS in four filters (B, V, R, I). In this paper we present the
photometry results for 50 radio galaxies. The observations were made in
1994--1998. A CCD ISD015A of $580\times520$ pixels with a  pixel size of
$0.205''\times0.154''$ was employed in the 1994--1995 observations, and a
CCD ISD017A ($1160\times1040, 0.137''\times0.137''$) was used in observations
of 1996--1998.
When reading the latter, the initial elements of the matrix were summed up
to produce the output elements twice as large in both coordinates. The exposure
 time was defined in the course of observations, proceeding
from brightness and colour of objects, from 1200--1800\,s (total) in the B
band to 400--800\,s in the R band for a typical object of our sample having
$R=22^m-23^m$, so that the resulting signal-to-noise ratio would be no worse
than 4--5 in all the filters. At a seeing worse than $2''$, we had to increase
the exposure time by a factor of 1.5--2.

Identifications of RC objects are presented in Fig.\,5 (see after the
references). The radio isophotes were plotted from the data of VLA observations
and superimposed on the images of galaxies obtained with the 6\,m telescope
in the R filter. The VLA radio observations were carried out in the 1990s
with different antenna configurations (Parijskij et al., 1996a). Besides,
the VLA data for the RC objects from the MIT survey (Fletcher et al., 1996)
and the FIRST survey (White et al., 1997) data were used. Images are also
presented for the objects that we classified as QSO and which are excluded
from the present study. The list of the objects is given in Table\,1.

\begin{table}
\caption{A list of RC objects classified as  QSO and excluded from this study}
\begin{center}
\begin{tabular}{|ccc|}
\hline
    Name      &  RA(2000.0) &  Dec(2000.0) \\
\hline
RC J0038+0449 & 00 38 34.65 & +04 50 50.5  \\
RC J0042+0504 & 00 42 27.17 & +05 05 24.7  \\
RC J0126+0502 & 01 26 16.11 & +05 02 09.9  \\
RC J0143+0505 & 01 43 33.89 & +05 07 57.4  \\
RC J0226+0512 & 02 26 19.80 & +04 46 32.5  \\
RC J0459+0456 & 04 59 04.28 & +04 55 54.4  \\
RC J0506+0558 & 05 06 25.00 & +05 08 19.3  \\
RC J1154+0431 & 11 54 53.50 & +04 24 12.5  \\
RC J1740+0502 & 17 40 33.96 & +05 02 42.3  \\
RC J2013+0508 & 20 13 23.48 & +05 10 30.5  \\
RC J2036+0451 & 20 36 56.93 & +04 49 52.7  \\
\hline
\end{tabular}
\end{center}
\end{table}

The processing of optical images was accomplished using the standard procedure
in the MIDAS system. Subtraction of the averaged dark frame and
element-by-element correction with the use of twilight sky frames were
performed. The
residual background inhomogeneity in the I filter, which is due to interference
of night sky emission lines, was removed by means of the procedure of
subtraction of the median sum of all working frames of a given night or
several nights of one observing run. The calibration of photometric
measurements for their conversion to the standard Johnson--Cousins system
was done using the stars from Landolt's (1992) list, which were observed
several times during a night. To do photometry of the selected object, a
circular aperture of one and the same size was used when measuring in
different filters. The aperture size was chosen to be between $3''$ and $12''$,
depending on the object luminosity. The typical size was $4''-5''$. The
background was measured with a circular aperture of sufficiently large
radius in order not to cover the outer regions of the object being measured.
When necessary, the neighbouring objects were removed by the procedure of
interpolation of the surrounding background. The photometry accuracy was
generally not worse than $0.1^m$ for  galaxies brighter than $21^m$. It dropped
down to $0.2^m-0.25^m$ for $23^m-24^m$ and was as low as $0.3^m-0.5^m$ for
galaxies fainter than $25^m-25.5^m$.

To provide for the Galactic extinction, we used the charts from the paper
``Maps of Dust IR Emission for Use in Estimation of Reddening  and
CMBR Foregrounds'' (Schlegel et al., 1998),
written in the form of FITS files. The conversion of stellar magnitudes to
flux densities was implemented via the formula
$S(Jy)= 10^{C - 0.4m}$ (von Hoerner, 1974).
The values of the constant $C$ for different filters are listed in Table\,2,
where the following characteristics are also given: name of filter,
effective wavelength, coefficient
$A/E(B-V)$ of change-over from distribution of dust radiation to extinction in
a given band under the assumption of extinction curve
$R_V=3.1$.

\begin{table}
\caption{Characteristics of photometric bands}
\begin{center}
\begin{tabular}{|l|r|c|c|}
\hline
  Filter name   &$\lambda_{eff}$&  $A/E(B-V)$  &   $C$      \\
\hline
  Landolt B     &      4400     &    4.315   &   3.620  \\
  Landolt V     &      5500     &    3.315   &   3.564  \\
  Landolt R     &      6500     &    2.673   &   3.487  \\
  Landolt I     &      8000     &    1.940   &   3.388  \\
\hline
\end{tabular}
\end{center}
\end{table}

The stellar magnitudes of 50 galaxies of our sample, which were
corrected for extinction in the four filters, are presented in Table\,3.

\begin{table*}
\caption{Stellar magnitudes of the sample radio galaxies corrected for
extinction}
\begin{center}
\begin{tabular}{|l|c|c|c|c||l|c|c|c|c|}
\hline
Source   &   B   &   V   &   R   &   I   & Source  & B  & V  & R  & I  \\
	   &   m   &   m   &   m   &   m   &           & m  & m  & m  & m  \\
\hline
0015+0503a & 23.89 & 22.97 & 22.20 & 21.36 & 1155+0444  & 21.36 & 19.83 & 18.90 & 18.20\\
0015+0501  & 24.82 & 23.91 & 23.37 & 22.22 & 1213+0500  & 23.55 & 22.90 & 22.04 & 21.32\\
0034+0513  & 25.28 & 24.79 & 23.25 & 21.79 & 1235+0435b & 24.15 & 22.81 & 21.59 & 20.35\\
0039+0454  & 24.81 & 24.00 & 22.69 & 21.22 & 1322+0449  & 23.68 & 22.52 & 20.77 & 19.17\\
0105+0501  & 24.00 & 22.48 & 22.78 & 22.43 & 1333+0452  & 24.87 & 24.44 & 23.56 & 22.46\\
0135+0450  & 20.49 & 19.16 & 18.42 & 17.82 & 1339+0445  & 25.05 & 23.72 & 22.70 & 21.55\\
0152+0453  & 23.31 & 23.02 & 22.47 & 21.70 & 1357+0453  & 22.98 & 21.85 & 21.10 & 20.11\\
0159+0448  & 22.65 & 21.72 & 21.23 & 20.65 & 1429+0501  & 25.57 & 23.24 & 21.64 & 20.50\\
0209+0501a & 20.37 & 19.19 & 18.43 & 17.78 & 1436+0501  & 23.90 & 23.86 & 23.39 & 22.70\\
0209+0501b & 25.72 & 24.09 & 23.12 & 21.63 & 1446+0507  & 21.48 & 20.03 & 19.17 & 18.54\\
0318+0456  & 25.64 & 23.98 & 22.61 & 20.99 & 1503+0456  & 24.02 & 23.67 & 23.14 & 22.24\\
0444+0501  & 23.48 & 23.70 & 23.33 & 23.13 & 1510+0438  & 24.98 & 23.73 & 22.57 & 21.25\\
0457+0452  & 22.01 & 20.86 & 20.05 & 19.37 & 1551+0458  & 25.57 & 25.34 & 24.43 & 23.30\\
0836+0511  & 23.68 & 23.53 & 23.09 & 22.44 & 1626+0448  & 22.32 & 23.07 & 22.73 & 22.63\\
0837+0446  & 23.03 & 23.29 & 22.99 & 22.11 & 1638+0450  & 22.86 & 22.33 & 22.14 & 21.04\\
0845+0444  & 24.72 & 22.42 & 21.09 & 19.77 & 1646+0501  & 24.01 & 22.44 & 20.97 & 19.76\\
0908+0451  & 21.63 & 20.72 & 19.85 & 19.07 & 1703+0502  & 24.22 & 23.39 & 23.12 & 22.26\\
0909+0445  & 22.60 & 21.53 & 20.50 & 19.59 & 1706+0502  & 24.73 & 24.19 & 23.25 & 21.88\\
0934+0505  & 25.29 & 24.45 & 24.67 & 23.61 & 1722+0442  & 22.30 & 21.59 & 20.63 & 19.44\\
1011+0502  & 23.71 & 23.18 & 22.47 & 22.60 & 2029+0456  & 22.85 & 22.24 & 21.66 & 20.53\\
1031+0443  & 23.93 & 22.79 & 22.09 & 20.85 & 2219+0458  & 24.80 & 25.03 & 23.72 & 22.25\\
1043+0443  & 23.98 & 23.57 & 22.51 & 21.70 & 2224+0513  & 23.16 & 22.31 & 21.43 & 20.32\\
1124+0456  & 20.30 & 18.79 & 17.85 & 17.07 & 2247+0507  & 23.64 & 23.18 & 22.53 & 21.43\\
1142+0455  & 24.83 & 22.53 & 21.38 & 20.39 & 2348+0507  & 23.89 & 23.79 & 23.56 & 23.08\\
1152+0449  & 23.86 & 23.66 & 22.39 & 21.00 &            &       &        &          &\\
\hline
\end{tabular}
\end{center}
\end{table*}

\begin{figure*}
\centerline{\psfig{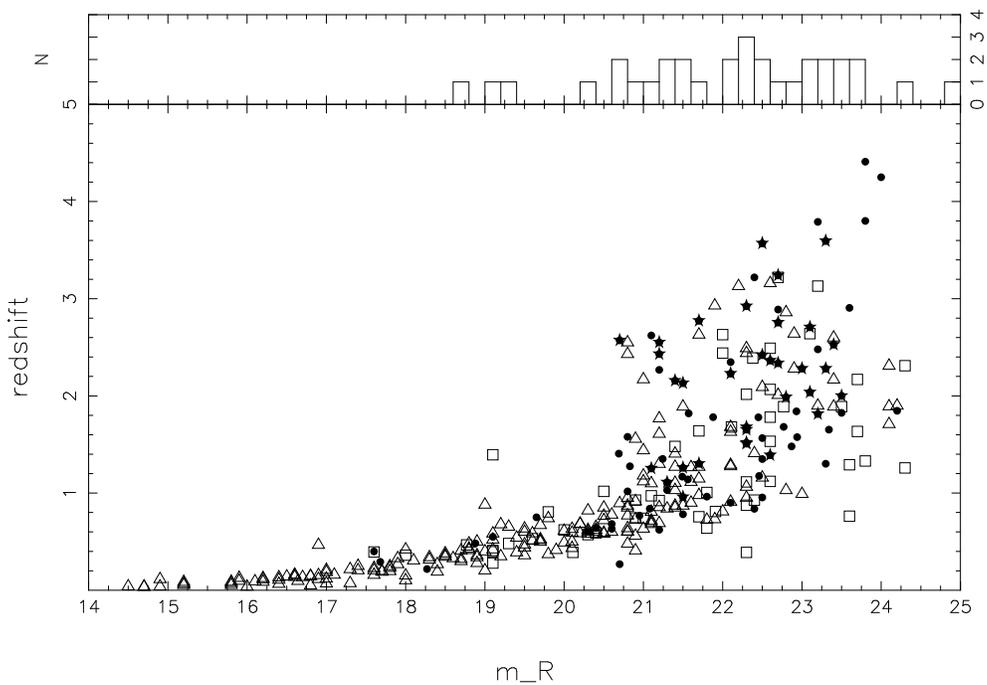}}
\caption{Hubble diagram R-value vrs redshift for different radio galaxies
taken from literature. The figure is reproduced from Pursimo et al. (1999).}
\end{figure*}

\section{Models of energy distribution in the spectra of host galaxies}
In the late 1980s and early 1990s, attempts were made to use colour
characteristics of radio galaxies for estimation of redshifts and ages of
systems of host galaxies. There appeared numerous evolution models with
which observational data were compared and results largely differing from
one another were obtained
(Arimoto \& Yoshii, 1987;
Chambers \& Charlot, 1990; Lilly,  1987, 1990; Parijskij et al., 1996a).
Over the past few years the models like  PEGASE:
Project de'Etude des Galaxies par Synthese Evolutive (Fioc \& Rocca-Volmerange,
1997), and GISSEL'98 (Bruzual, Charlot, 1993; Bolzonella et al., 2000), in
which an attempt was made to eliminate the shortcomings of the preceding
versions, have been most commonly used.

In the ``Big Trio'' experiment
(Parijskij et al., 1996a), we also tried to apply these methods to distant
steep spectrum objects of the RC catalogue, for which we measured the
magnitudes in the four filters (BVRI). This paper is distinguished from the
paper of Parijskij et al. (1996a) by using more reliable identifications of
radio sources, new photometric data and new models of energy distribution.
Besides, we used in our work the smoothing procedure that makes it possible
to model and predict the flux in the given filter at the specified SED with
allowance for the filter throughput curve and also for the redshift effect.
These changes in the procedure enabled making the results of estimation
more reliable as compared to our previous work.

We (Verkhodanov et al., 1999) preliminarily discussed applicability of the
new methods to a population of distant ($z>1$) radio galaxies with known
redshifts, for which we managed to find in literature more or less reliable
data of multicolour photometry in the optical and near infrared ranges in
no fewer than three filters. It is shown, in particular, that redshifts can be
estimated with an accuracy of 25--30\,\% at $1<z<4$, given measured stellar
magnitudes in more than three filters. And given at least one luminosity
estimate in the infrared region, it suffices to use measurement in three
filters. Estimates were made for two evolution models.

As one of the SED models the evolution model
PEGASE (Fioc \&
Rocca-Volmerange, 1997) was employed for galaxies of the Hubble sequence both
star forming and evolving in a passive fashion. One of the principal merits
of this model is the extension to the near infrared range (NIR) of the
atlas of synthetic spectra of
Rocca-Volmerange \& Guiderdoni (1988)
with the revised stellar library including parameters of cool stars. The
model covers a range from 220\,\AA~ to 5 microns. According to the authors
the algorithm of the model makes it possible to follow rapid evolution
phases such as red supergiants or AGB in the near IR. We used a wide set of
SED curves from this model in a range of ages from
$7 \times 10^6$ years to $19 \times 10^9$ years for massive elliptical
galaxies.

From the library of synthetic spectra of the model
GISSEL'98
(Bolzonella et al., 2000)
derived with the aid of the evolution models of
Bruzual \&  Charlot (1993, 1996),
we used the computations for elliptical galaxies. The library of synthetic
spectra was constructed with the following star formation parameters:
simple stellar population (SSP), the duration of starburst activity is
1 billion years, the starburst activity decays exponentially.
The solar metallicity was used in the model. The initial mass function (IMF)
with an upper limit of 125 solar masses was taken from
Miller \& Scalo (1979). As Bolzonella et al. (2000) show the choice of IMF
has no effect on the accuracy of redshift measurements. The mode tracks are
computed in a wavelength range from 200 to 95800\,\AA. We used the range
specified by a redshift limit of 0 to 6 in our calculations.

The sets of evolution models employed are available at
{\it http://sed.sao.ru} (Verkhodanov et al., 2000).

\section{Estimation of age and redshift}
Before using the model transmission curves we performed their smoothing
with the filter
via applying the algorithm
$$
  S_{ik} = \frac {\sum\limits_{j=0}^n s_{i-n/2+j} f_{jk}(z)}
		 {\sum\limits_{j=0}^n f_{jk}(z)},
$$
where
$s_i$ is the initial model curve of SED,
$S_{ik}$ is the model curve of SED smoothed by the $k$th filter,
$f_k(z)$ is the curve of transmission of the $k$th filter ``compressed''
$(1+z)$ times when ``moving'' along the wavelength axis of the SED curve,
$j=1, n$ is the number of the pixel in the filter transmission curve.
From the $k$  curves of SED thus formed (there are four of them in our case),
we constructed a two-dimensional array
($\lambda$, filter) of smoothed spectra for further computations.

The evaluation of ages and redshifts was carried out by the method of choosing
an ``optimum'' location of photometric magnitudes obtained in photometric
observations of radio galaxies in different filters on the SED curves.
We used the already computed and table-specified SED curves for different
ages. The algorithm of selection of the optimum location on the curve was
briefly as follows (for details see Verkhodanov, 1996): by way of shifting
the points lengthwise (along the wavelength axis) and transverse (along the
intensity axis) the SED curves, we defined the location at which the sum of
squares of deviations of points from the corresponding smoothed curves is a
minimum, i.e. the minimum of $\chi^2$ is actually calculated:
$$
\chi^2 = \sum\limits_{k=1}^{N filters}
      \left( \frac{F_{obs,k} - p\cdot {\tt SED}_{k}(z)}{\sigma_k}\right),
$$
where $F_{obs,k}$ is the observed stellar magnitude in the $k$th filter,
${\tt SED}_{k}(z)$ is the model stellar magnitude for the given spectral
distribution in the $k$th filter at a given $z$,
$p$ is the free coefficient, $\sigma_k$ is the measurement error.

The redshift was found from the shift of location of the observed magnitudes
at their best location on the SED curves from the ``rest frame'' position
(position at $z=0$). From the total set of curves for different ages we
chose the ones on which the sum of squares of deviations for the given
observations turned out to be a minimum.

Fig.\,6 (see at the end of the paper) gives the results of the most
plausible choice of evolution models for the photometry data and the
corresponding function of plausibility.
\vspace{-0.3cm}

\section{Photometric redshifts for RC catalogue radio galaxies}

To examine the potentialities of the method for determination of redshift
and ages of the stellar population of host galaxies from the data of
multicolour photometry, we selected 40 remote radio galaxies with known
redshifts, for which stellar magnitudes in no fewer
than three filters are available in literature
 (Verkhodanov et al., 1998b, 1999).

It should be noted that literature photometric data are most inhomogeneous:
they were obtained by different authors, different tools, with
different filters; measurements for one and the same object were not always
made in the same apertures, etc. This is why, only 42 out of 300 radio
galaxies of the original sample remained. The subsamples of objects
(3C, 4C, B2) with relatively homogeneous data are not large enough to be
statistically compared. At first, using the collected photometric data
obtained with the aid of the models PEGASE and GISSEL'98, only the ages
of the stellar population of host galaxies  with a fixed known
redshift were derived. After that, a search for an optimum model SED curve,
for the redshift and age of the stellar population simultaneously, was
carried out, and  the derived values were compared.

Thus, we estimated both the galaxy age and the redshift within the frames
of the specified models (see Verkhodanov et al., 1998a, 1999). It is clear
from general considerations that the reliability of the result at large
redshifts shows a significant dependence on the availability of infrared
data (up to the R range), because when fitting, we ``cover'' the region of
rapid spectrum change (a jump) before the optical range of SED and thereby
can reliably (with a well defined maximum of the plausibility curve)
determine the location of our data. This can be proved as follows. When the
available points are removed (to check the reliability of the procedure)
with retaining only 3 points (one of which is in the K range), we obtain
in fitting the same result on the curve of deviations as for 4 or 5 points.
If the infrared range is not used, the result will then be more uncertain.
However, as we show (Verkhodanov et al., 1999) the version of ``tight''
positioning of the four filters, as in our case of BVRI photometry, gave a
good result in a sample of 6 galaxies, which is a good fit to the result
obtained with using all the filters, including the infrared range.

A similar procedure was employed to estimate the redshift and the age for
galaxies of the RC catalogue. The common distinction from the papers by
Verkhodanov et al., 1998b, 1999) is that our observations were made only in
4 filters, B\,V\,R\,I, so, the infrared data are lacking.

Table\,4 presents redshift and age estimates for the stellar systems of radio
galaxies from the RC catalogue.
The redshift  $z_c$, the age of the radio galaxy
 $t$, the discrepancy $\varepsilon$ and the age of the Universe $T$ at $z=z_c$
($H=65, \Omega_{m}=0.3, \Omega_{\Lambda}=0.7$ for
standard Friedmann-Lema\^itre-Robertson-Walker cosmology
(Thomas \& Kantowski, 2000)) are given in the table columns for the models
GISSEL and PEGASE, respectively.

\begin{center}
\begin{onecolumn}
\vspace*{0.5cm}
\topcaption{Colour redshifts and ages of stellar systems for the
 RC Catalogue radio galaxies}
\tablefirsthead{ \hline
     \multicolumn{1}{|c|}{}
   & \multicolumn{4}{|c|}{}
   & \multicolumn{4}{|c|}{} \\
     \multicolumn{1}{|c|}{Object RC J}
   & \multicolumn{4}{|c|}{GISSEL'98}
   & \multicolumn{4}{|c|}{PEGASE}  \\
     \cline{2-9}   & & & & & & & & \\
     \multicolumn{1}{|c|}{}                     
   & \multicolumn{1}{|c|}{$z_c$}                
   & \multicolumn{1}{|c|}{t[Myr]}               
   & \multicolumn{1}{|c|}{$\varepsilon$}        
   & \multicolumn{1}{|c|}{T[Myr]}               
   & \multicolumn{1}{|c|}{$z_c$}                
   & \multicolumn{1}{|c|}{t[Myr]}               
   & \multicolumn{1}{|c|}{$\varepsilon$}        
   & \multicolumn{1}{|c|}{T[Myr]} \\            
\hline
     \multicolumn{1}{|c|}{1}                    
   & \multicolumn{1}{|c|}{2}                    
   & \multicolumn{1}{|c|}{3}                    
   & \multicolumn{1}{|c|}{4}                    
   & \multicolumn{1}{|c|}{5}                    
   & \multicolumn{1}{|c|}{6}                    
   & \multicolumn{1}{|c|}{7}                    
   & \multicolumn{1}{|c|}{8}                    
   & \multicolumn{1}{|c|}{9} \\                 
\hline
}
\tablehead{ \hline
     \multicolumn{1}{|c|}{1}                    
   & \multicolumn{1}{|c|}{2}                    
   & \multicolumn{1}{|c|}{3}                    
   & \multicolumn{1}{|c|}{4}                    
   & \multicolumn{1}{|c|}{5}                    
   & \multicolumn{1}{|c|}{6}                    
   & \multicolumn{1}{|c|}{7}                    
   & \multicolumn{1}{|c|}{8}                    
   & \multicolumn{1}{|c|}{9} \\                 
\hline
}
\tabletail{ \hline }
\begin{supertabular}{|l|r|r|r|r|r|r|r|r|}
\hline
0015+0503a   & 0.73 &    900& 0.0412 &   7546 &  0.48 &  3000& 0.0299&  9237 \\
	     & 0.07 &  16000& 0.0342 &  13512 &  4.91 & 16000& 0.0199&  1273 \\
0015+0501    & 0.81 &   1000& 0.0334 &   7103 &  0.87 &  3250& 0.0358&  6797 \\
0034+0513    & 0.98 &  16000& 0.0729 &   6286 &  1.29 & 10000& 0.0374&  5130 \\
	     &      &       &        &        &  1.11 & 15000& 0.0196&  5755 \\
0039+0454    & 0.95 &  16000& 0.0346 &   6419 &  0.99 &  7000& 0.0057&  6242 \\
0105+0501    & 0.28 &    500& 0.0852 &  11032 &  4.21 &  9250& 0.0839&  1537 \\
0135+0450    & 0.35 &   2500& 0.0106 &  10350 &  0.10 &  7250& 0.0018& 13112 \\
	     & 0.12 &  16000& 0.0072 &  12855 &       &      &       &       \\
0152+0453    & 0.75 &    400& 0.0051 &   7431 &  0.81 &   640& 0.0016&  7103 \\
	     & 1.15 &    900& 0.0113 &   5606 &       &      &       &       \\
0159+0448    & 0.13 &   2000& 0.0093 &  12729 &  0.41 &  1900& 0.0093&  9813 \\
	     &      &       &        &        &  0.09 &  5000& 0.0090& 13244 \\
	     &      &       &        &        &  4.75 & 15000& 0.0060&  1326 \\
0209+0501a   & 0.39 &   1800& 0.0067 &   9988 &  0.39 &  3500& 0.0063&  9988 \\
	     &      &       &        &        &  0.09 &  6750& 0.0088& 13244 \\
	     &      &       &        &        &  0.07 &  9250& 0.0097& 13512 \\
	     &      &       &        &        &  0.07 & 16000& 0.0058& 13512 \\
0209+0601b   & 0.69 &  13000& 0.0400 &   7783 &  0.70 &  6000& 0.0483&  7723 \\
0318+0456    & 0.67 &  16000& 0.0491 &   7906 &  0.71 &  6500& 0.0071&  7663 \\
0444+0501    & 1.17 &    200& 0.0365 &   5534 &  2.42 & 13000& 0.0419&  2870 \\
0457+0452    & 0.40 &   2000& 0.0092 &   9900 &  0.41 &  3500& 0.0068&  9813 \\
0836+0511    & 0.77 &    250& 0.0058 &   7319 &  0.81 &   286& 0.0094&  7103 \\
	     & 1.14 &    600& 0.0028 &   5643 &  1.21 &  2100& 0.0037&  5394 \\
	     & 1.93 &   2500& 0.0229 &   3599 &  1.37 &  2400& 0.0021&  4886 \\
0837+0446    & 0.98 &    200& 0.0490 &   6286 &  0.99 &   202& 0.0494&  6242 \\
	     & 1.91 &   6000& 0.0697 &   3636 &  1.92 &  9250& 0.0635&  3617 \\
0845+044     & 5.29 &  16000& 0.1139 &   1160 &  0.66 &  8750& 0.0560&  7969 \\
0908+0451    & 0.49 &   1400& 0.0322 &   9159 &  0.48 &  3250& 0.0193&  9237 \\
	     &      &       &        &        &  4.99 & 16000& 0.0125&  1248 \\
0909+0445    & 0.67 &   1800& 0.0145 &   7906 &  0.64 &  4000& 0.0168&  8096 \\
0934+0505    & 1.72 &   1200& 0.0785 &   4010 &  1.79 &  4250& 0.0728&  3865 \\
	     &      &       &        &        &  1.79 & 16000& 0.0727&  3865 \\
1011+0502    & 0.46 &    300& 0.0501 &   9397 &  0.50 &   453& 0.0624&  9082 \\
1031+0443    & 0.86 &   2500& 0.0188 &   6846 &  0.87 &  4250& 0.0275&  6797 \\
1043+0443    & 1.19 &   3000& 0.0138 &   5463 &  0.72 &  2200& 0.0250&  7604 \\
	     &      &       &        &        &  1.23 &  4500& 0.0171&  5326 \\
	     &      &       &        &        &  5.98 &  8500& 0.0181&   993 \\
1124+0456    & 0.35 &  10000& 0.0025 &  10350 &  0.35 &  5750& 0.0115& 10350 \\
1142+0455    & 5.20 &  16000& 0.1082 &   1185 &  4.99 &   571& 0.0386&  1248 \\
	     &      &       &        &        &  0.35 & 14000& 0.0421& 10350 \\
1152+0449    & 1.17 &  16000& 0.0619 &   5534 &  1.33 &  8000& 0.0069&  5006 \\
	     &      &       &        &        &  1.30 & 15250& 0.0079&  5099 \\
1155+0444    & 4.18 &    500& 0.1161 &   1550 &  0.33 &  5750& 0.0179& 10538 \\
	     & 0.35 &   9000& 0.0091 &  10350 &       &      &       &       \\
1213+0500    & 0.70 &    700& 0.0201 &   7723 &  0.68 &  2100& 0.0162&  7844 \\
	     & 1.08 &   2000& 0.0334 &   5871 &  5.15 & 15500& 0.0118&  1200 \\
	     & 0.07 &   9000& 0.0522 &  13512 &       &      &       &       \\
1235+0435b   & 0.64 &  11000& 0.0042 &   8096 &  0.68 &  5750& 0.0083&  7844 \\
1322+0449    & 0.65 &  16000& 0.1129 &   8032 &  0.78 &  7250& 0.0702&  7264 \\
	     &      &       &        &        &  0.99 & 14000& 0.0037&  6242 \\
1333+0452    & 1.04 &   1400& 0.0158 &   6032 &  1.08 &  4000& 0.0062&  5871 \\
1339+0445    & 0.67 &   4500& 0.0217 &   7906 &  0.71 &  5000& 0.0136&  7663 \\
1357+0453    & 0.73 &   1400& 0.0421 &   7546 &  0.41 &  4000& 0.0409&  9813 \\
1429+0501    & 5.38 &    450& 0.0827 &   1135 &  0.61 & 11500& 0.0483&  8294 \\
1436+0501    & 1.35 &    800& 0.0157 &   4945 &  1.15 &  1680& 0.0136&  5606 \\
	     & 1.93 &   3000& 0.0318 &   3599 &  1.18 &  1900& 0.0127&  5499 \\
1446+0507    & 0.35 &   5000& 0.0140 &  10350 &  0.24 &  5750& 0.0191& 11451 \\
	     &      &       &        &        &  4.99 &  6750& 0.0189&  1248 \\
	     &      &       &        &        &  0.11 & 14500& 0.0144& 12983 \\
1503+0456    & 0.78 &    500& 0.0065 &   7264 &  0.83 &  1015& 0.0023&  6998 \\
1510+0438    & 0.66 &   8000& 0.0110 &   7969 &  0.75 &  5750& 0.0074&  7431 \\
1551+0458    & 1.05 &   1200& 0.0341 &   5991 &  1.11 &  4000& 0.0294&  5755 \\
	     & 1.32 &   6000& 0.0099 &   5036 &  1.32 &  5750& 0.0129&  5036 \\
1626+0448    & 0.04 &    200& 0.1268 &  13930 &  0.03 &   202& 0.1282& 14073 \\
	     & 2.31 &   6000& 0.1642 &   3011 &  2.30 &  3000& 0.1814&  3025 \\
1638+0450    & 0.84 &    500& 0.0315 &   6947 &  0.89 &  1015& 0.0388&  6699 \\
	     & 1.67 &   2000& 0.0284 &   4119 &  1.78 &  4750& 0.0134&  3885 \\
1646+0501    & 0.65 &$>$5000& 0.1171 &   8032 &  0.64 &  6750& 0.0106&  8096 \\
1703+0502    & 0.81 &    600& 0.0692 &   7103 &  3.56 &   640& 0.0567&  1874 \\
	     & 0.07 &   2500& 0.0431 &  13512 &  0.35 &  1680& 0.0508& 10350 \\
	     & 0.00 &   6000& 0.0452 &  14515 &  0.00 &  5500& 0.0363& 14515 \\
1706+0502    & 0.96 &   1800& 0.0339 &   6374 &  1.06 &  5750& 0.0293&  5951 \\
	     & 1.17 &  16000& 0.0317 &   5534 &       &      &       &       \\
1722+0442    & 1.02 &   4000& 0.0033 &   6115 &  1.01 &  4500& 0.0033&  6157 \\
2029+0456    & 0.81 &    800& 0.0094 &   7103 &  0.88 &  2600& 0.0107&  6747 \\
2219+0458    & 0.98 &   1200& 0.1271 &   6286 &  1.36 &  7750& 0.0555&  4916 \\
	     & 1.22 &  16000& 0.0987 &   5360 &       &      &       &       \\
2224+0513    & 0.73 &   1200& 0.0101 &   7546 &  0.77 &  3500& 0.0072&  7319 \\
2247+0507    & 0.78 &    700& 0.0094 &   7264 &  0.93 &  2500& 0.0068&  6510 \\
	     & 1.38 &   3000& 0.0253 &   4857 &  1.45 &  4750& 0.0134&  4660 \\
2348+0507    & 1.43 &    450& 0.0016 &   4715 &  1.24 &   571& 0.0030&  5293 \\
	     &      &       &        &        &  2.07 &  3000& 0.0291&  3362 \\
\hline
\end{supertabular}
\end{onecolumn}
\end{center}

\section*{Notes to Table\,4}
{\bf 0034+0513} --- the age of the stellar system in the two alternatives
from the model GISSEL'98 is greater than that of the Universe at the given
	     $z$; we choose the age where the difference is smaller despite
a somewhat greater error. \\
{\bf 0039+0454} --- the age of the stellar system from the PEGASE model is
inadmissively large, it is about 3 times as large as that of the Universe
at a given $z$. The same is observed in a number of other sources
(0209+0501,
	     0845+0454, 1142+0455,
	     1152+0449, 1235+0435b, 1322+0449, 1510+0438,
	     1646+0501,
	     2219+0454).
The cause of this has not been understood yet.\\
{\bf 0105+0501} --- in contrast to the majority of objects of our sample the
redshift and the age from the two models are not consistent. Besides, the
computation gives large errors in this case. The optical images (Soboleva
et al., 2000) show that the bright line $Ly\alpha$ falls in the V filter.
For this reason, the stellar magnitude is strongly corrupted. \\
{\bf 0135+0450} --- the redshift value lies between 0.34 and 0.10, but
closer to 0.34. Note that from the GISSEL'98 model there is another minimum
at  $z=0.17$ with the age 5750 Myr and a deviation of
	     0.0105. The values at the absolute minimum are discarded in
the two models because of the discrepancy between the age of the system and
that of the Universe.\\
{\bf 0159+0448} --- $z=0.09$ (GISSEL'98) and $z=0.13$ (PEGASE) lie outside
the region of permissible values in the plane $z - R$ (see Fig.\,1;
	     Pursimo et al., 1999).
	     The age 15000 Myr in the  GISSEL model does not conform to
reality. We consider the alternative with
	     $z=0.41$ to be more reliable.\\
{\bf 0209+0501a} --- $z=0.39$ for both models. The age derived does not
conform to real fact for a minimum deviation 0.0058 in the GISSEL model.
Two other estimates yield a larger value of deviation. \\
{\bf 0318+0456} --- close redshifts for the models GISSEL'98 ($z=0.71$)
	     and PEGASE ($z=0.67$). The age turns out to be ultimately
small for both models.\\
{\bf 0444+0501} --- although to the value  $z=1.17$ for the PEGASE model
corresponds a minimum deviation, the closest to the spectroscopic redshift
	     ($z=2.73$) is obtained in the
 GISSEL model ($z=2.42$). The estimated age proves to be too great to be
real, although $z$ is close to real.\\
{\bf 0836+0511} --- the values  $z=1.21$ (GISSEL'98) and $z=1.14$ (PEGASE)
are close (to an accuracy of 6\,\%) and practically have a minimal deviation.
One can see here that the age of stellar systems turns out to be older in the
GISSEL models.\\
{\bf 0845+0444} --- $z=5.29$ (PEGASE) is not admissible both by age and
by position in the plane  $z - R$.\\
{\bf 0908+0451} --- the colour redshift ($z_{ph}=0.48$) is confirmed by spectral
measurements at the 6 m telescope of SAO RAS
($z_{sp}=0.5$).
	     \\
{\bf 1142+9455} --- $z=4.99$ (GISSEL'98) and $z=5.20$ (PEGASE)
do not fall within the range of admissible values in the plane
$z - R$.  We choose $z=0.35$, though in this case the age of the stellar
system, according to the criterion of minimum of squares of deviations, proves
to be somewhat larger than the age of the Universe.\\
{\bf 1213+0500} --- $z=5.15$ (GISSEL) is outside of the region of acceptable
values in the plane $z - R$, and the age of the stellar system in this case
is 10 times that of the Universe. There are close redshift values in the
region  $z=0.7$ for both models, a minimum $\chi^2$ being consistent with
this solution in the PEGASE model.          \\
{\bf 1322+0449} --- the variant with  $z=0.99$ in the  GISSEL model is
unsuitable because the age of the stellar system is twice that of the Universe.
A similar situation also occurs in the PEGASE model where the solution goes
beyond the limits of acceptable age values. There is a stable minimum for
 $z=0.78$ in the PEGASE model. In all the cases the redshift proves to be
less than 1.\\
{\bf 1357+0453} --- assume $z=0.73$ from  PEGASE model.
	     $z=0.41$, however, for  GISSEL model falls out of the plane
$z-R$. Note that there is one more local minimum for the  GISSEL  model,
which corresponds to  $z=3.86$ in the plane of distribution of deviations, but
the value of this minimum is 3 times as large as for
$z=0.41$, and the corresponding point also falls out of the plane
$z-R$. \\
{\bf 1429+0501} --- we fail to find acceptable values of the parameters
for either model. This may be due to difficulties of separating a radio
galaxy from a nearby star in the optical range (Parijskij et
	     al., 1996a).\\
{\bf 1436+0501} --- accept the alternatives $z=1.18$ (GISSEL) and
	     $z=1.35$ (PEGASE) with errors close to minimum. The difference
between the redshifts in these models is then about 12\,\%.\\
{\bf 1446+0507} --- at $z=0.11$ the age of the galaxy in the  GISSEL model
turns out to be larger than the age of the Universe at a minimum of deviations,
at $z=4.99$ with a minimum error it does not fall within the range of
admissible values in the plane  $z - R$ for radio galaxies,
and the age of the stellar system is 5 times the age of the Universe.
We retain  $z=0.35$ for PEGASE and $z=0.24$ for GISSEL as possible alternatives.\\
{\bf 1626+0448} --- spectroscopic redshift for this object is
 $z=2.66$ (Afanasiev  et al., 2002). The current calculations have been
carried out with no provision for the line $Ly\,{\alpha}$. The data obtained
show that  $z=0.03$ and $z=0.04$ do not fall in the interval of acceptable
values for radio galaxies in the plane  $z - R$. In all the cases the large
errors are due to the fact that the bright line $Ly\,{\alpha}$  falls on
the B band  (Parijskij et al., 1996a).
The  $z$ values equal to  2.31 for PEGASE and 2.30 for GISSEL
are quite close to the true values. When the line  $Ly\,{\alpha}$
is taken into account in the computations, the redshift value then turns out
to be equal to the spectroscopic.\\
{\bf 1638+0450} --- choose $z=1.78$ (GISSEL'98) and $z=1.67$
	    (PEGASE),
though $z=0.89$ (GISSEL'98) and $z=0.84$ (PEGASE)  cannot be rejected either. \\
{\bf 1703+0502} --- the stellar magnitudes in all the filters can be affected
by a nearby bright star (Parijskij et al., 1996a).\\
{\bf 1722+0442} --- the spectroscopic $z=0.7$ (Afanasiev et al., 2002)
is by 30\,\% lower than the value derived by photometry for both models.\\
{\bf 2247+0507} --- there are two minima in the two models.
The $z$ values less than 1 differ for the models by less than 20\,\%.
	     The difference is less than 10\,\% for  $z$ of about  1.4,
however, the age of the stellar system is the same as the age of the Universe
for the model  GISSEL. By the criterion $\chi^2$ we choose the former
alternative in both models.\\
{\bf 2348+0507} --- the stellar magnitudes in all the filters can be affected
by a nearby bright star (Parijskij et al., 1996a).

\section{Discussion and results}
{\bf 1. The problem of colour redshifts of distant galaxies}

As it was shown earlier (Verkhodanov et al., 1999), the model PEGASE yielded
satisfactory results over 40 objects for which spectral measurements of
redshifts are available. Analysis of Table 4 shows that the new model
CISSEL often gives results only slightly differing from the PEGASE model
(see also Verkhodanov et al., 2001a, b).

\begin{figure}
\centerline{\psfig{figure=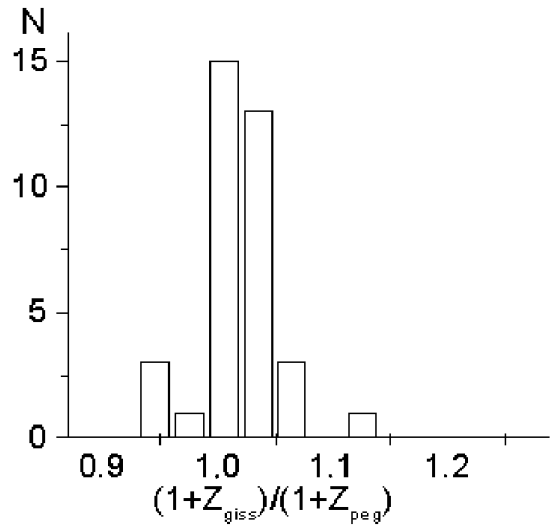,width=8cm}}
\caption{Histogram characterizing the distribution of the ratio of colour
redshifts from the GISSEL'98  ($z_{giss}$) and  PEGASE ($z_{peg}$) models.
The central part of the histogram, within which  80\,\% of objects fall,
is presented.}
\end{figure}

Fig. 2 presents a histogram of the ratio $(1+z_{GISSEL})/(1+z_{PEGASE})$.
The distribution of errors is not normal, there is a nucleus with very small
model errors, in which most of the objects (80\,\%) fall, but 20\,\% of the objects
have very large errors. As a rule, the large errors are caused by getting of the
strong line ($ Ly\,{\alpha}$)  in one of the filters, illumination from a
nearby bright object and complicated cases of SED deviation from the model.
A number of such examples are described in notes to Table 4.

It should be noted that we use the standard set of filters of the 6 m telescope,
BVRI, which is satisfactory for not too distant objects (the colours of which
have been measured yet). Measurements have to be extended to the H and
K ranges. In a considerable number of cases, this will enable elimination
of uncertainties in the estimates. The situation may change for ultimately
distant objects with R$>24^m$, and we have to wait until the observations are
completed, prior to giving recommendations. It is clear from general
considerations that there is a danger from ``right'' and ``left'', that is,
the secondary star formation sites may distort the blue region of the spectrum, but
the dust at very great redshifts may deform the IR region. It can easily be shown
that large discrepancies may point to possibilities of luminosity distortion
by strong lines, and we hope to take this into account in the future.
For  $z>2$, it is necessary to take into account, at least, the line
$Ly\,{\alpha}$.  Considerable errors may arise because of the periodicity
of series in Bor's model, but here simple energetic considerations may be
helpful (discrepancies in the colour and photometric estimates). This can be
illustrated by the RC object J1703+0502 formally having a zero redshift.
When adopting this estimate, the optical luminosity of this object will
then prove to be so low that it cannot produce any perceptible radio radiation
(note that $P_{radio} {\sim} L_{opt}^{2.5}$;
Iskudarian  \& Parijskij, 1964; Franceschini  et  al., 1998). When making
use of the secondary criteria, we sometimes have to reject the variant
with the smallest discrepancies and take the next one.

{\bf 2. The age of stellar systems of host galaxies}

As it is known, the age is much more difficult to assess than the redshift.
The older the stellar population the larger the error may be. We have made
a histogram (Fig. 3) of differences of ages determined from the models
GISSEL ($t_{giss}$) and  PEGASE ($t_{peg}$)  for one and the same object. They
are normalized to the age of the Universe with ${\Lambda}=0.7$ ($T$) for the
moment that corresponds to the measured redshift $(t_{giss}-t_{peg})/2T$.
\begin{figure}
\centerline{\psfig{figure=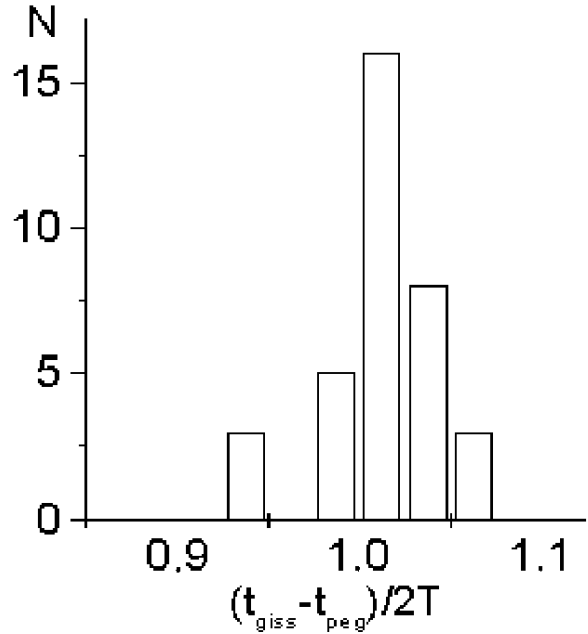,width=8cm}}
\caption{Histogram of half-difference of colour ages of host stellar systems
obtained from the models GISSEL'98 $(t_{giss})$ and PEGASE $(t_{peg})$  vrs the
age of the Universe with $\Lambda = 0.7$ $(T)$ for the moment corresponding
to the measured redshift. The central part of the histogram is shown where
80\,\% of objects fall.}
\end{figure}
The distribution of differences of ages derived from  different models
are far from normal, but there is a nucleus where 70\,\% of objects fall.
The ``model'' dispersion for them is close to 20\,\%. Large departures
were detected in 9 out of 50 objects with the use of PEGASE. The measured
age of host stellar systems for them is far larger than that of the Universe
even at redshifts close to GISSEL. For this reason, we consider the GISSEL
model to be more trustworthy. A systematic difference between the models
has been discovered, which increases with rising redshift (see Fig. 4).
When the difference is taken into account, the dispersion of ages drops to
$<10$\,\% for objects with close redshifts in both models.

\begin{figure}
\centerline{\psfig{figure=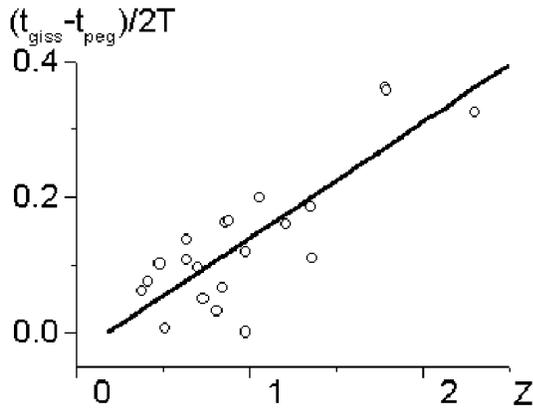,width=8cm,angle=90}}
\caption{Systematic difference between ages of stellar systems from GISSEL'98
$(t_{giss})$ and
PEGASE $(t_{peg})$ models normalized to the age of the Universe with $\Lambda = 0.7$
 $(T)$ for the moment of the corresponding redshift as a function of redshift.
The departures from the relationship derived are 6\,\%.}
\end{figure}

It is essential that not a single GISSEL model galaxy  is older within the errors
than the Universe in the model with the ${\Lambda}$-term equal to 0.7--0.8.
There are more than ten such galaxies in the old model SCDM, and they can be
used for independent assessment of the behaviour of $H(z)$, ${\Lambda}$-term
and quintessence (suggestion of  Starobinsky (1998), see
Saini et al. (2000); Parijskij et al. (1998)). The average age of galaxies is
a few billion years, however, there is a subgroup of galaxies having the age
that coincides within the errors with the age of the Universe at a
corresponding redshift, and a group of formally very young (some hundred
million years) stellar systems. The formers are likely to belong to the
first generation of galaxies in the Universe which formed at
$z\gg 1$ ($z\sim 5-10$). The situation of the galaxies with ``young'' objects is
more complicated, their colour can be distorted by repeated star-formation bursts
at merging of galaxies or under the action of close passings. The variant of
``young'' galaxy cannot be accepted, since we deal with powerful radio galaxies,
for functioning of which a supermassive black hole is necessary
(DMO --- Dark Massive Object,  Salucci et al., 1999) with the mass of about
$10^9 M_\odot$, that is not possible within the standard models of black hole
formation (Franceschini et al., 1998). Only the primary black holes with masses
$10^4-10^6 M_\odot$, round of which galaxies form later on, may be the
alternatives of ``merging''.

As a result, a careful conclusion can be drawn that at least the statistical
estimates of the redshift and age for population of powerful radio galaxies
give satisfactory results. The GISSEL model can be recommended not only for
radio-quiet galaxies but also for powerful radio galaxies. It is advantageous
to use all available data on the objects to obtain more reliable estimates, this
will decrease the errors.

In conclusion we enumerate the results obtained.
\begin{enumerate}
\item
The largest currently available body of data on BVRI magnitudes of host
galaxies responsible for the origin of powerful radio galaxies is presented.
\item
The colour redshifts for powerful radio galaxies show a satisfactory agreement
with the spectral ones (the error is 10--20\,\% with a small fraction of
large errors). The recent 2001 spectral observations   of the RC objects
RC\,J0908+0451,  RC\,J1154+0431, RC\,J1626+0448 É RC\,J1722+0442 gave errors
of redshift measurement by the techniques described above within 10--15\,\%
(Afanasiev et al., 2002).
\item
The limited number of closely spaced filters as in our BVRI case may also
yield satisfactory
results even for large redshifts.
\item
The redshift distribution for the studied objects of our sample (the
subgroup of objects  brighter than $m_R=23.5^m$) shows a maximum near
$z{\sim} 1$, i.e. in the range of maximum radio activity of the Universe. The
group of objects with large colour redshifts $(z > 2.5-3)$ requires a
separate analysis. In any case, we do not consider yet the gap in the population
of the region  $(1.5 < z < 2.5)$  to be real.
\item
The colour data are generally not at variance with the stellar magnitudes
in the filter R (Parijskij et al., 1996a) when $R<22.5^m$. Fainter objects
show a higher dispersion of photometric redshifts at one and the same value
of R. One can notice here two branches in the plane $(z-R)$ (Pursimo et al., 1999).
The search for differences in the morphology, radio luminosity, spectral
indices has given no final result. Note that the objects with the steepest
spectra and a large radio-to-optical luminosity ratio occur in the branch
with large redshifts. However, further investigations are needed. Objects
with low relative radio luminosity, as it was to be expected, prove to be
either quasars or nearby galaxies.
\item
The age of galaxies is determined less reliably, and results turn out to
be of low significance for larger $z$. However, in practice one can
always indicate the lower limit of the age of galaxies and, therefore,
the minimum redshift of their formation. This age is always larger than the
standard estimate of the object lifetime, and in a number of cases it exceeds
the age of the Universe at an object redshift in a simple CDM model. There
are no such galaxies in a model with a ${\Lambda}$-term of 0.6--0.8
(e.g. de Bernardis et al., 2000).
\item
A programme has been developed and tested of automatic determination of
colour redshifts and ages of galaxies, which is applicable to any redshifts
with provision for transformation of the shape of the filters when changing
over from the rest to the moving system of reading, available through the
server {\it http://sed.sao.ru}.
\end{enumerate}
We contemplate further investigation of colour and photometric techniques
as applied to the population of distant radio galaxies. New more realistic
models of colour evolution and refined methods of age evaluation in stellar
systems will make it possible to obtain more reliable results for a great
number of objects.

\begin{acknowledgements}
The work was done with partial support from the SSTP ``Astronomy''
(projects 1.2.1.2 É 1.2.2.4),
through the grants of RFBR 99-02-17114, INTAS 97-1192, Integration project
 578, 206.
OVV, AIK and OPZh thank RFBR (grant 99-07-90334) for support in the creation
of a server for calculation of redshifts and ages of radio galaxies
({\it http://sed.sao.ru}). Thanks are due to the editor-in-chief, N.F.
Vojkhanskaya, and also to the referee for remarks, recommendations and
discussions.
\end{acknowledgements}

\clearpage
\newpage

\setcounter{page}{17}

\begin{figure*}
\begin{center}
Fig. 5. Oprical identifications of RC sources: VLA radio isophotes
overlaid on the 6\,m telescope R-band images, and R-band isolines
overalaid on the VLA images.\vspace*{0.5cm}
\centerline{
\vbox{
\psfig{figure=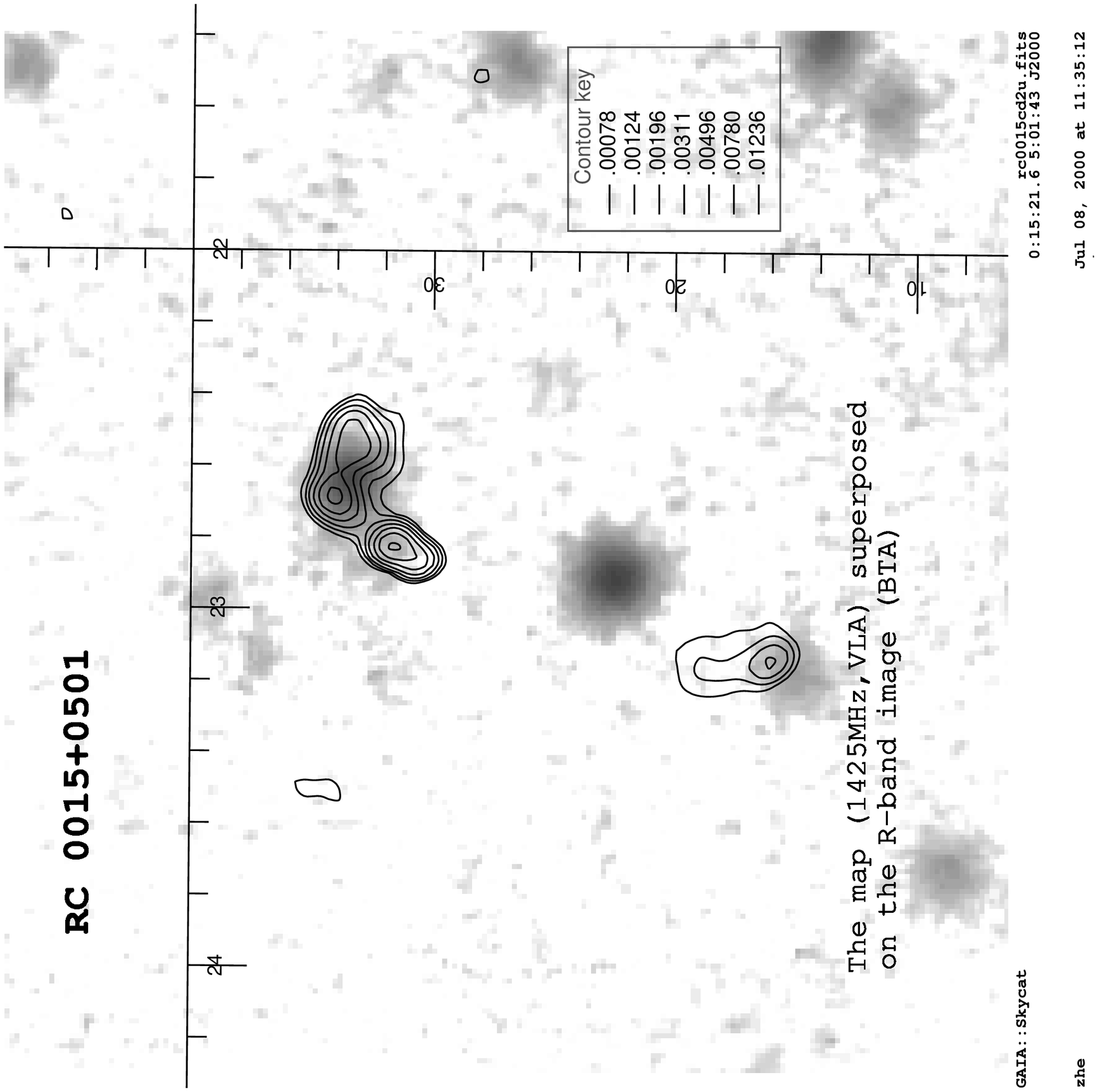,angle=-90,width=11cm}
}}
\end{center}
\end{figure*}
\begin{figure*}
\centerline{\psfig{figure=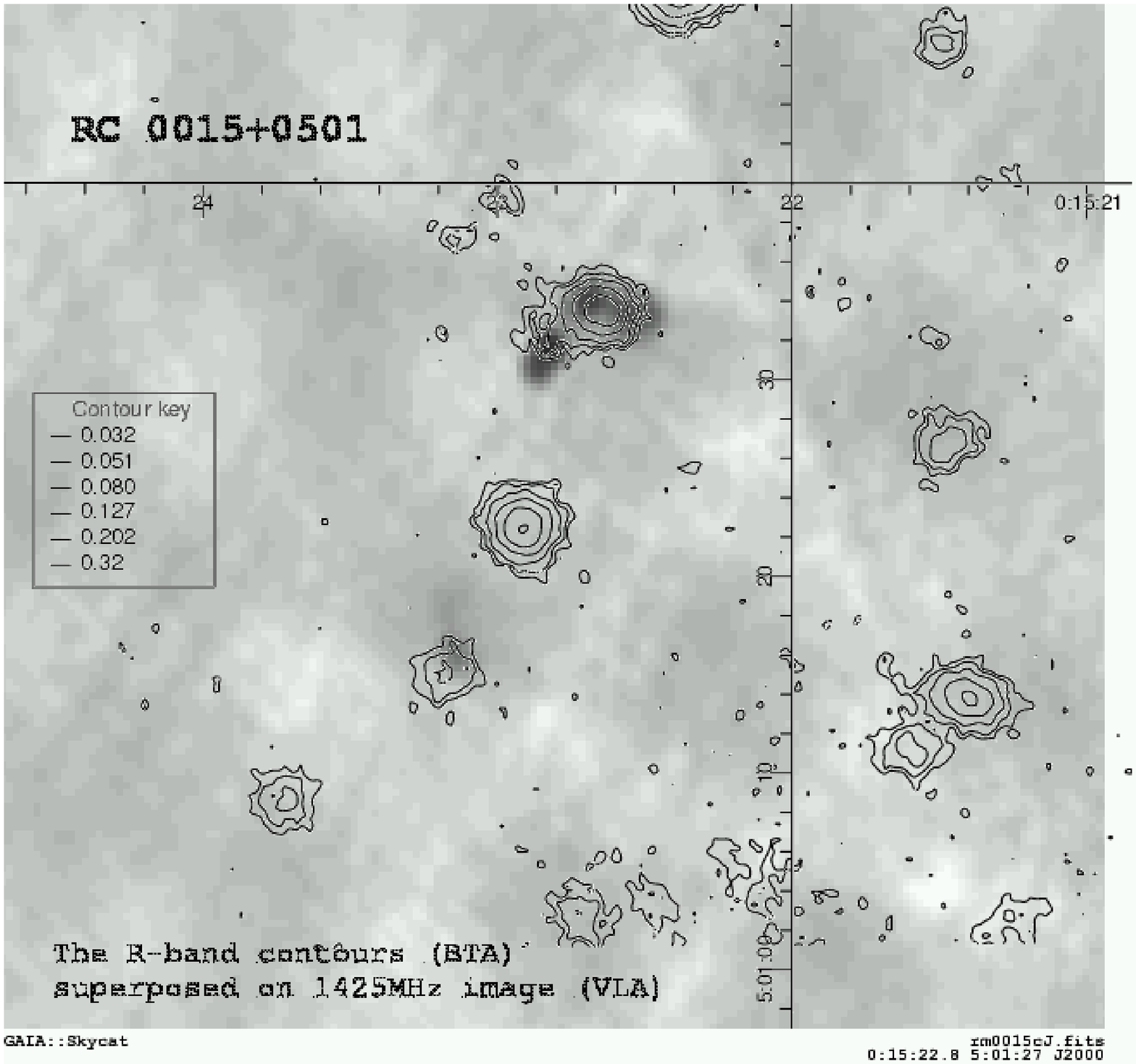,width=11cm}}
\end{figure*}

\begin{figure*}
\centerline{\psfig{figure=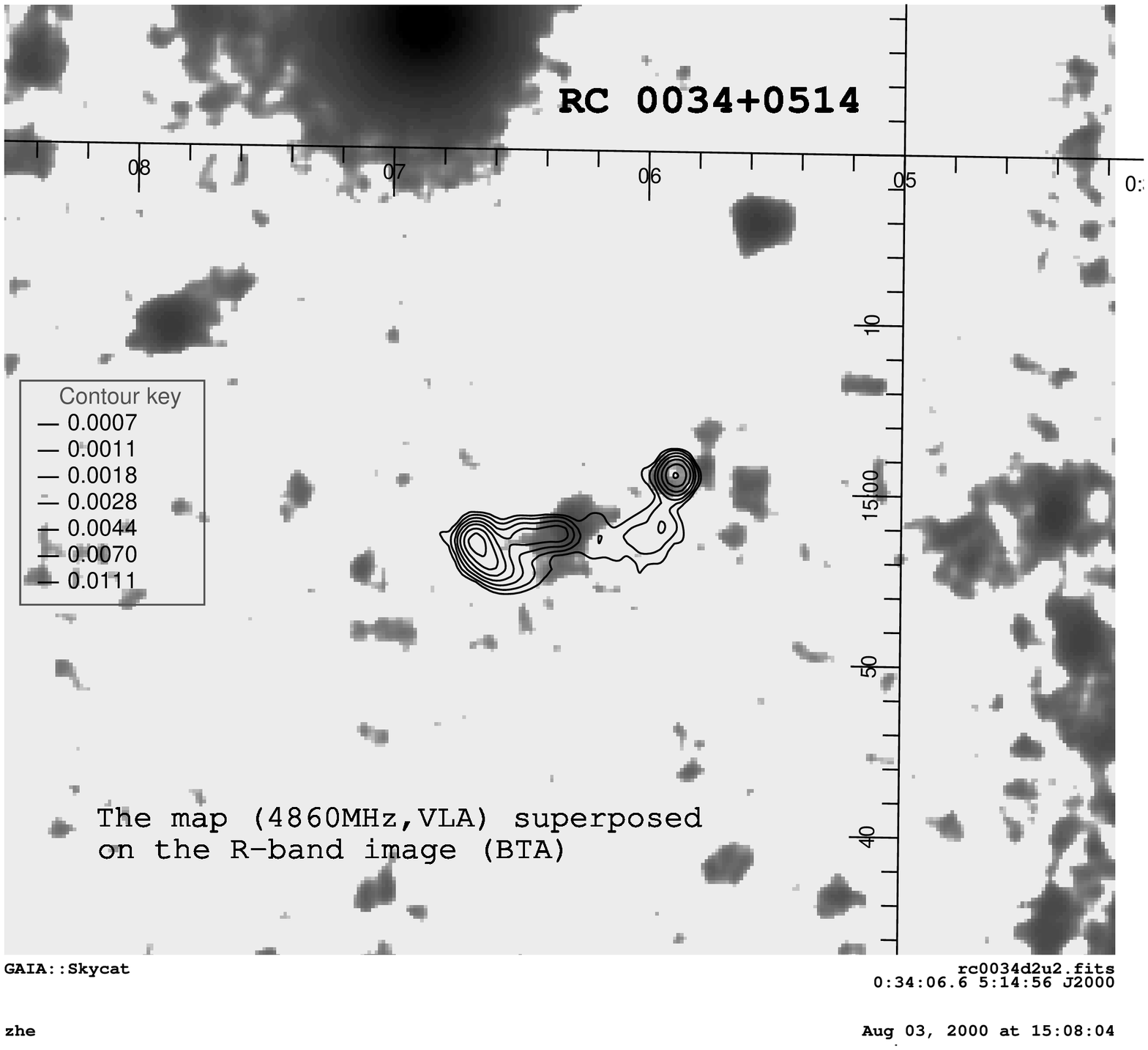,angle=-90,width=11cm}}
\end{figure*}
\begin{figure*}
\centerline{\psfig{figure=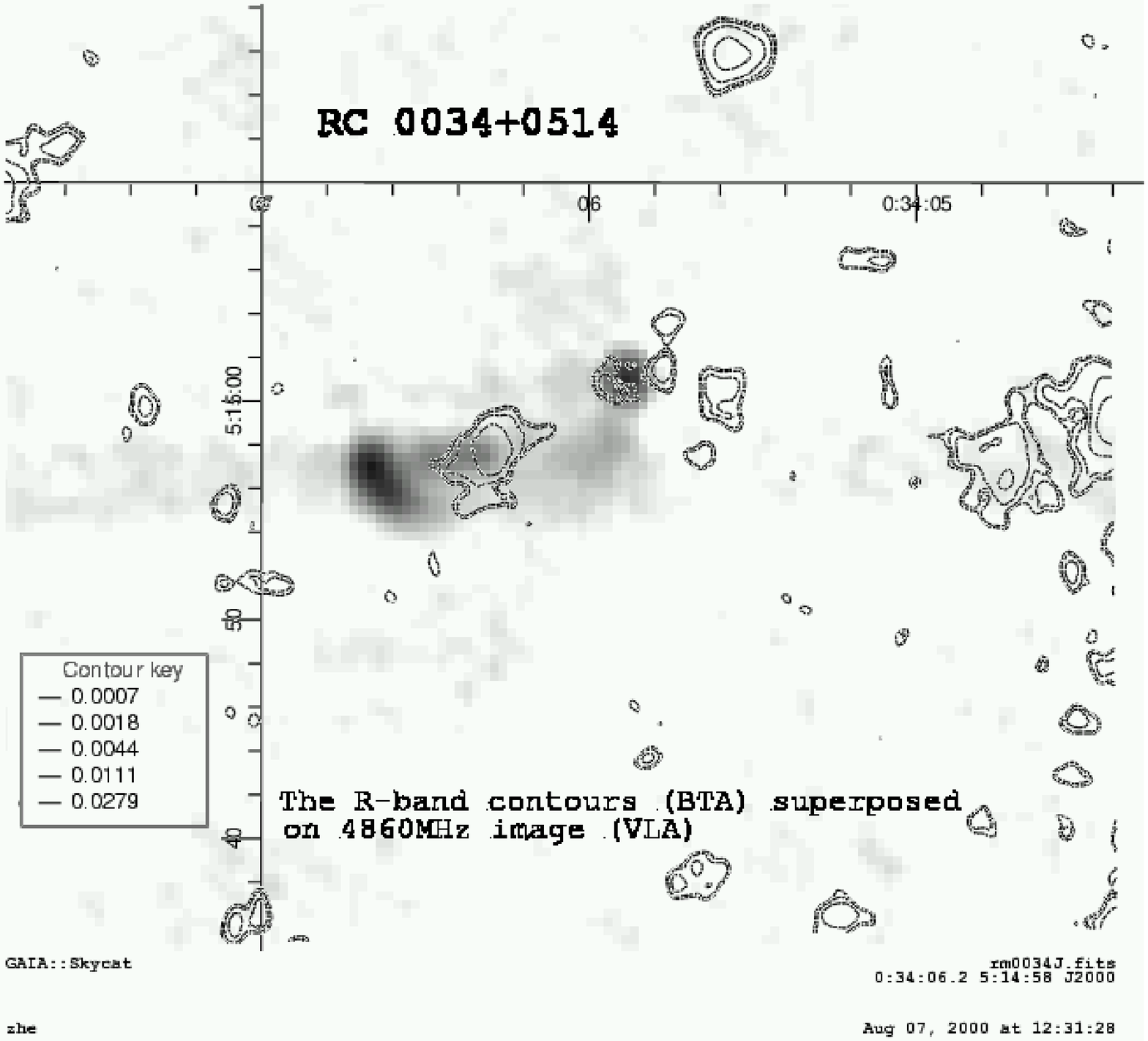,width=12cm}}
\end{figure*}

\clearpage
\newpage

\setcounter{page}{90}

\begin{figure*}[!h]
\vspace*{-1.0cm}
Fig. 6. Fitted SED models and corresponding probability functions for RC objects.
\vspace*{0.5cm}
\begin{center}
\vbox{
\hbox{
\mbox{\hspace*{0.7cm}}
\psfig{figure=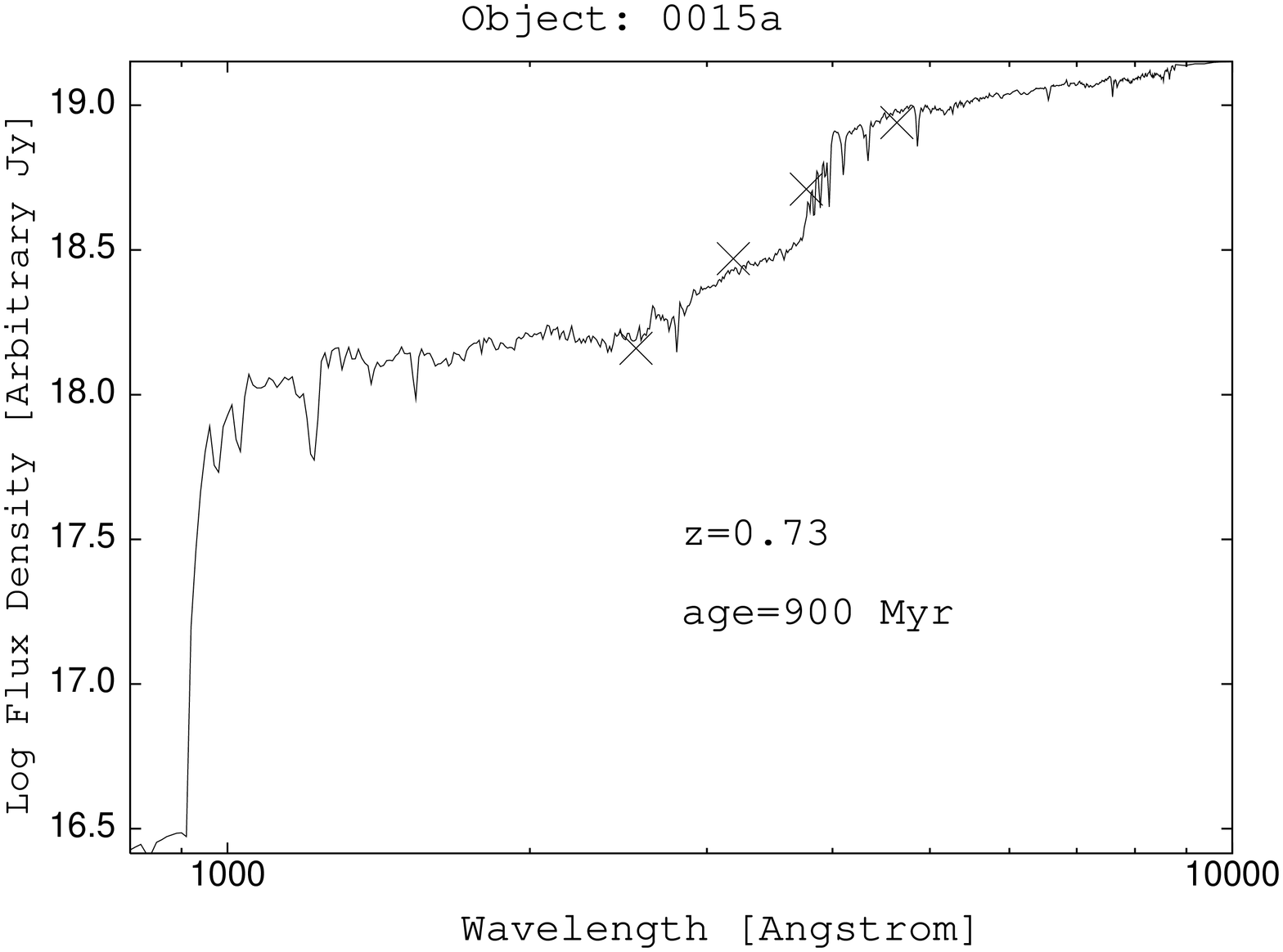,width=7cm,angle=-90}       
\mbox{\hspace*{1cm}}
\psfig{figure=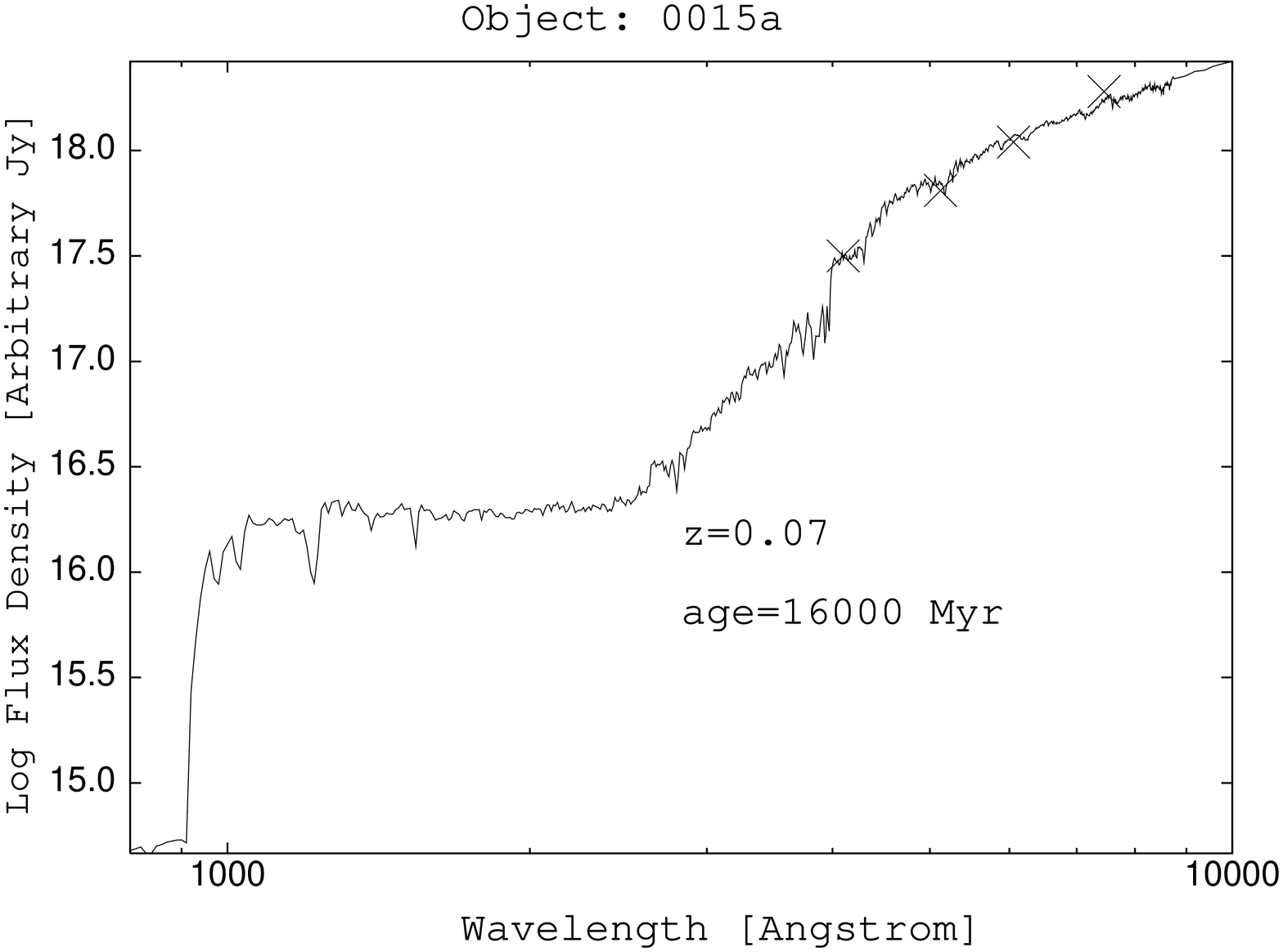,width=7cm,angle=-90}     
}
\hbox{
\mbox{\hspace*{0.1cm}}
\psfig{figure=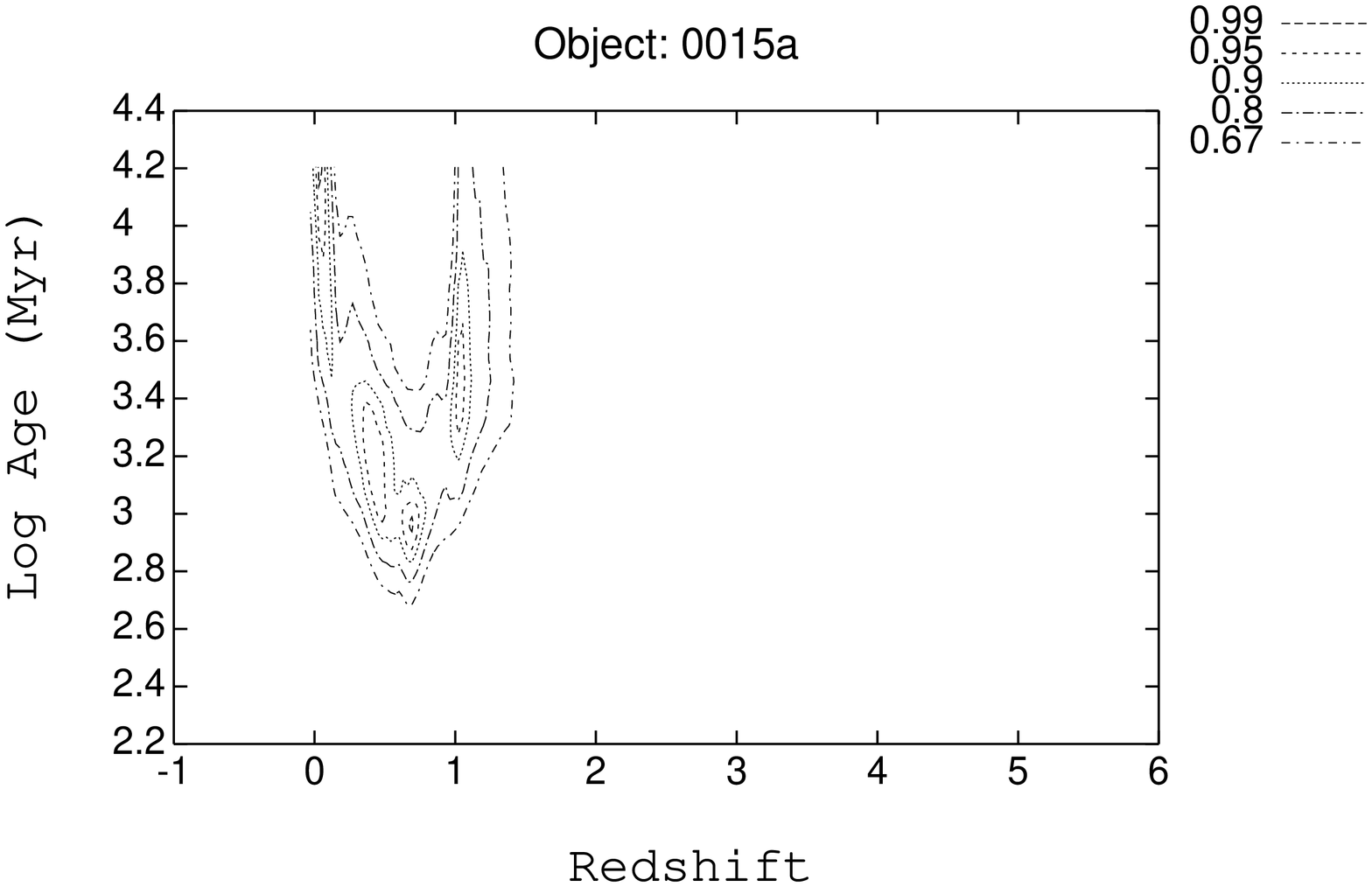,width=8cm,angle=-90,bbllx=520pt,bblly=90pt,bburx=68pt,bblly=745pt,clip=}
\mbox{\hspace*{10cm}}
}
\mbox{PEGASE models for RC J0015+0503a: SEDs and probability function}
\vbox{
\mbox{\vspace*{2cm}}
}
\hbox{
\mbox{\hspace*{0.7cm}}
\psfig{figure=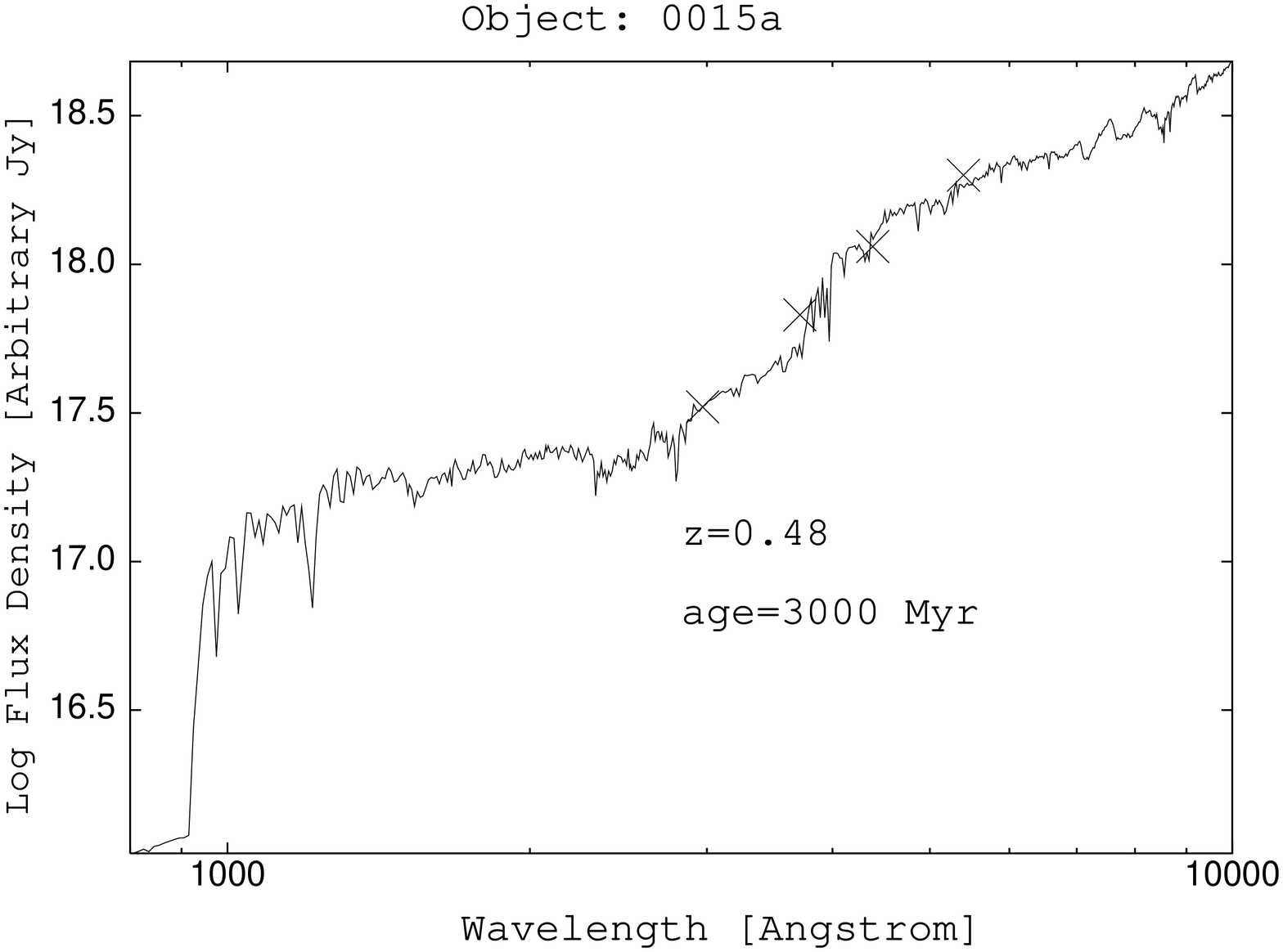,width=7cm,angle=-90}       
\mbox{\hspace*{1cm}}
\psfig{figure=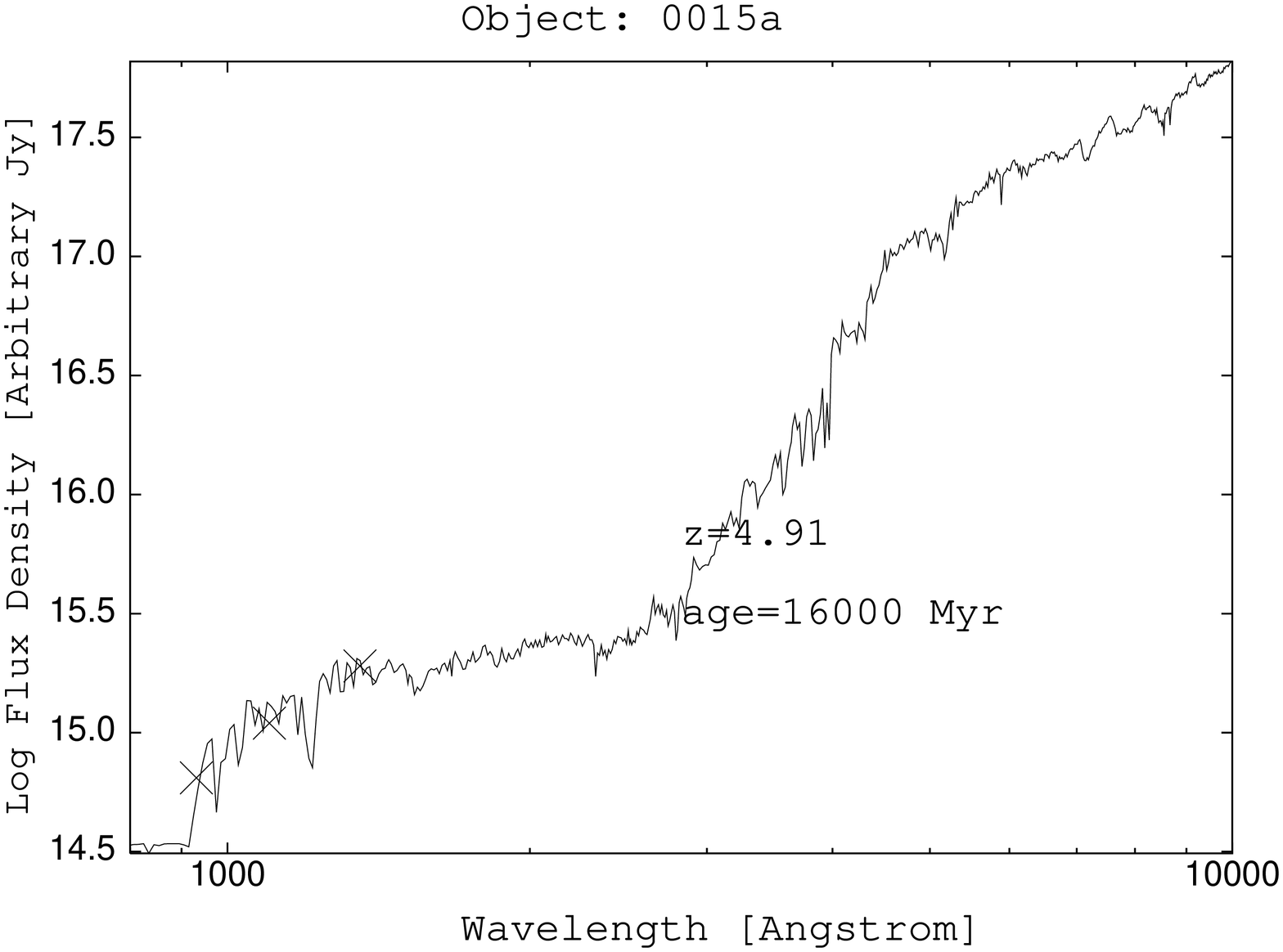,width=7cm,angle=-90}      
}
\hbox{
\mbox{\hspace*{0.2cm}}
\psfig{figure=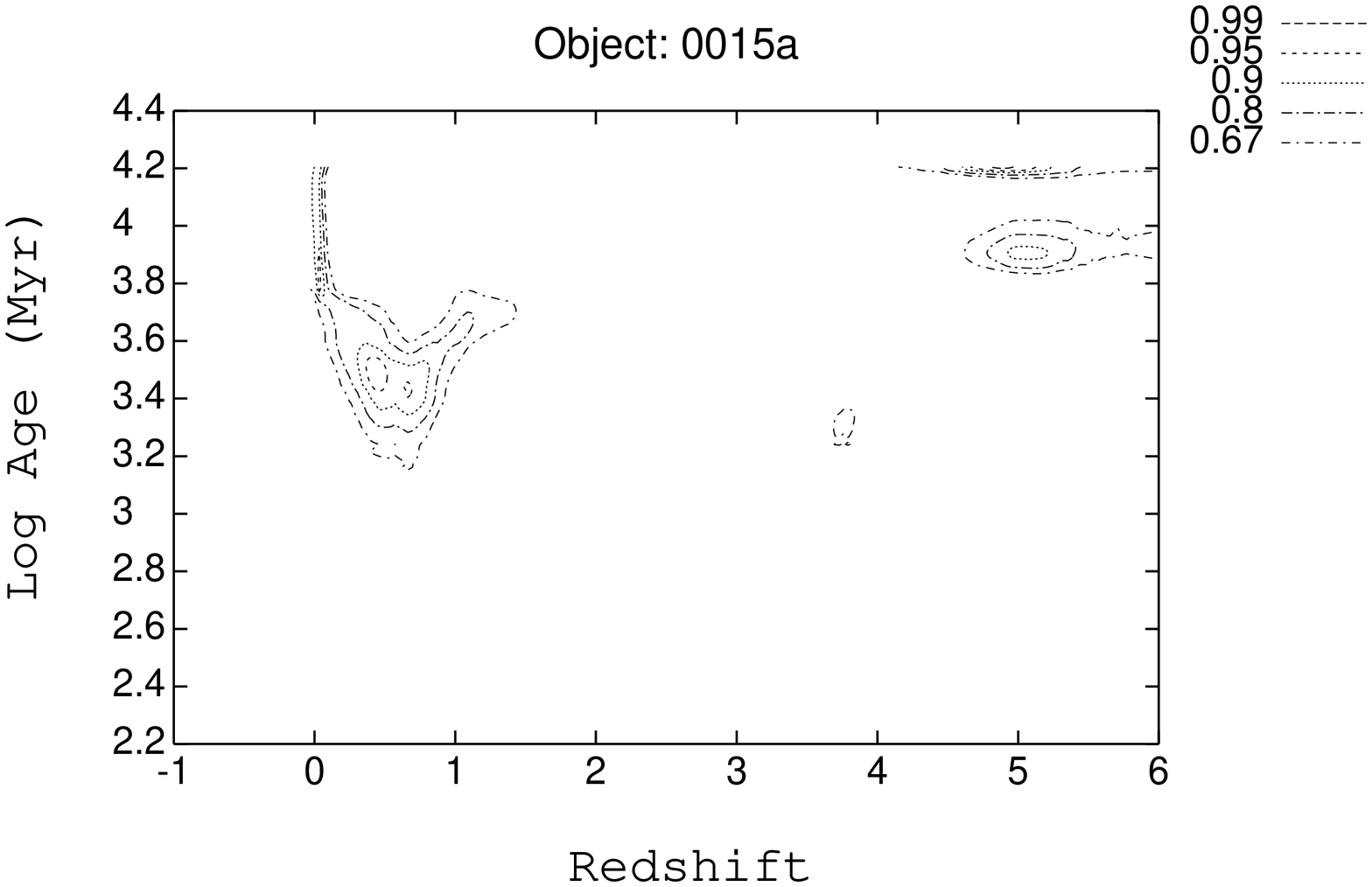,width=8cm,angle=-90}
\mbox{\hspace*{10cm}}
}
\mbox{GISSEL models for RC J0015+0503a: SEDs and probability function}
}
\end{center}
\end{figure*}

\newpage
\clearpage

\begin{figure*}[!h]
\centerline{PEGASE models for RC J0015+0501} 
\vspace*{-0.5cm}
\centerline{
\hbox{
\mbox{\hspace*{0.5cm}}
\psfig{figure=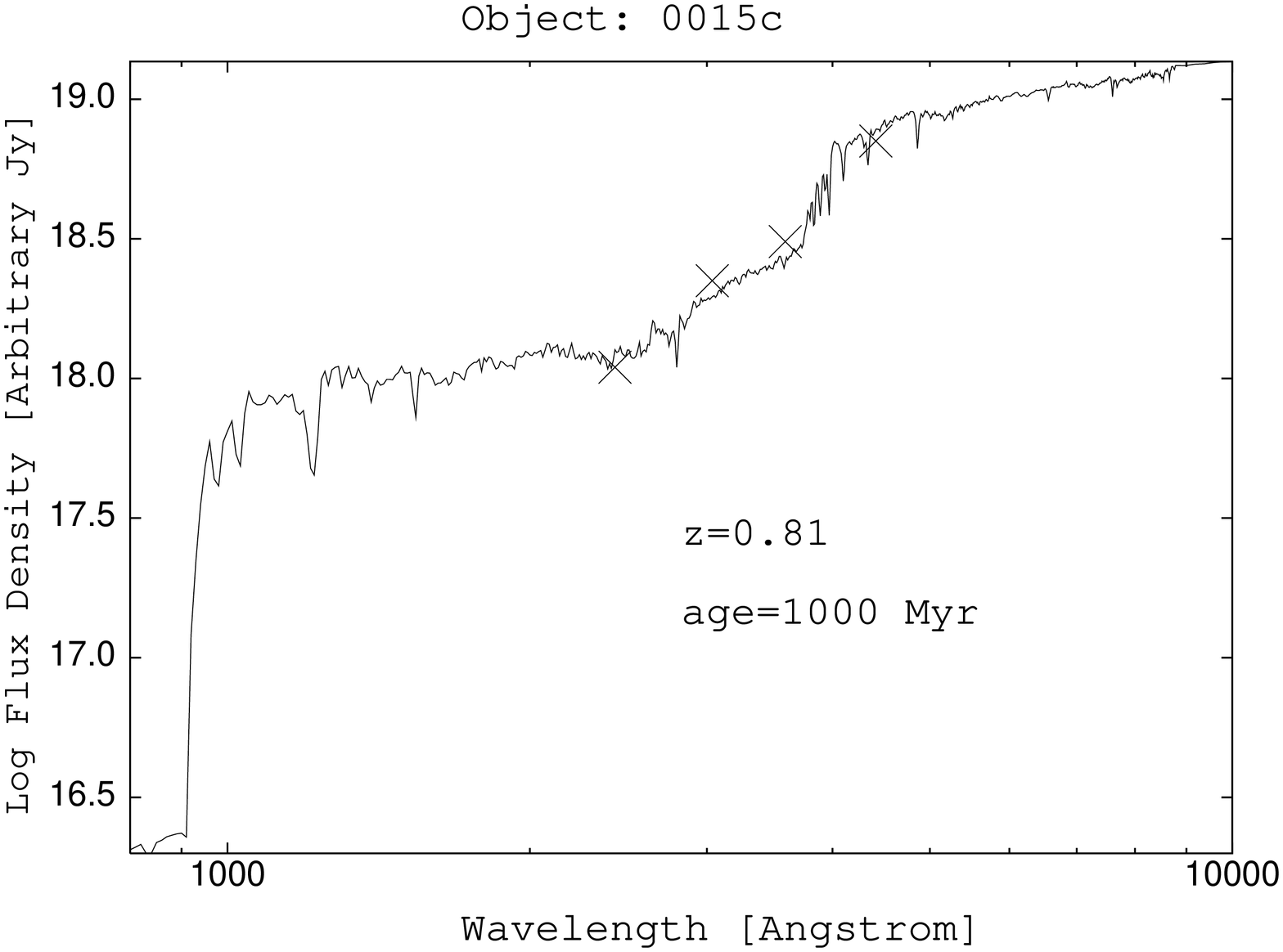,width=8cm,angle=-90}             
\psfig{figure=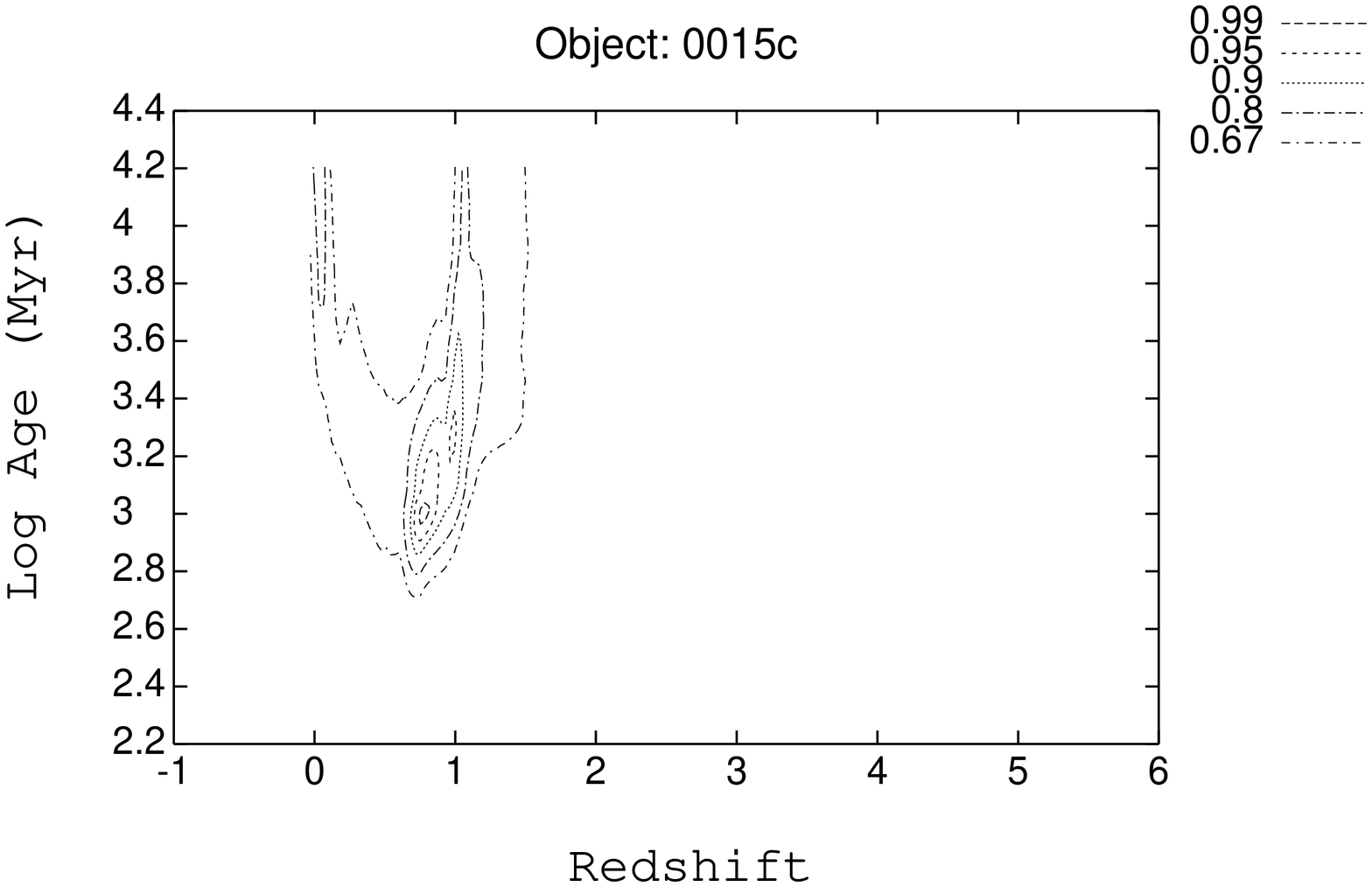,width=9cm,angle=-90,bbllx=520pt,bblly=90pt,bburx=68pt,bblly=745pt,clip=}
}}
\end{figure*}

\begin{figure*}[!h]
\centerline{GISSEL models for RC J0015+0501} 
\centerline{
\hbox{
\mbox{\hspace*{0.5cm}}
\psfig{figure=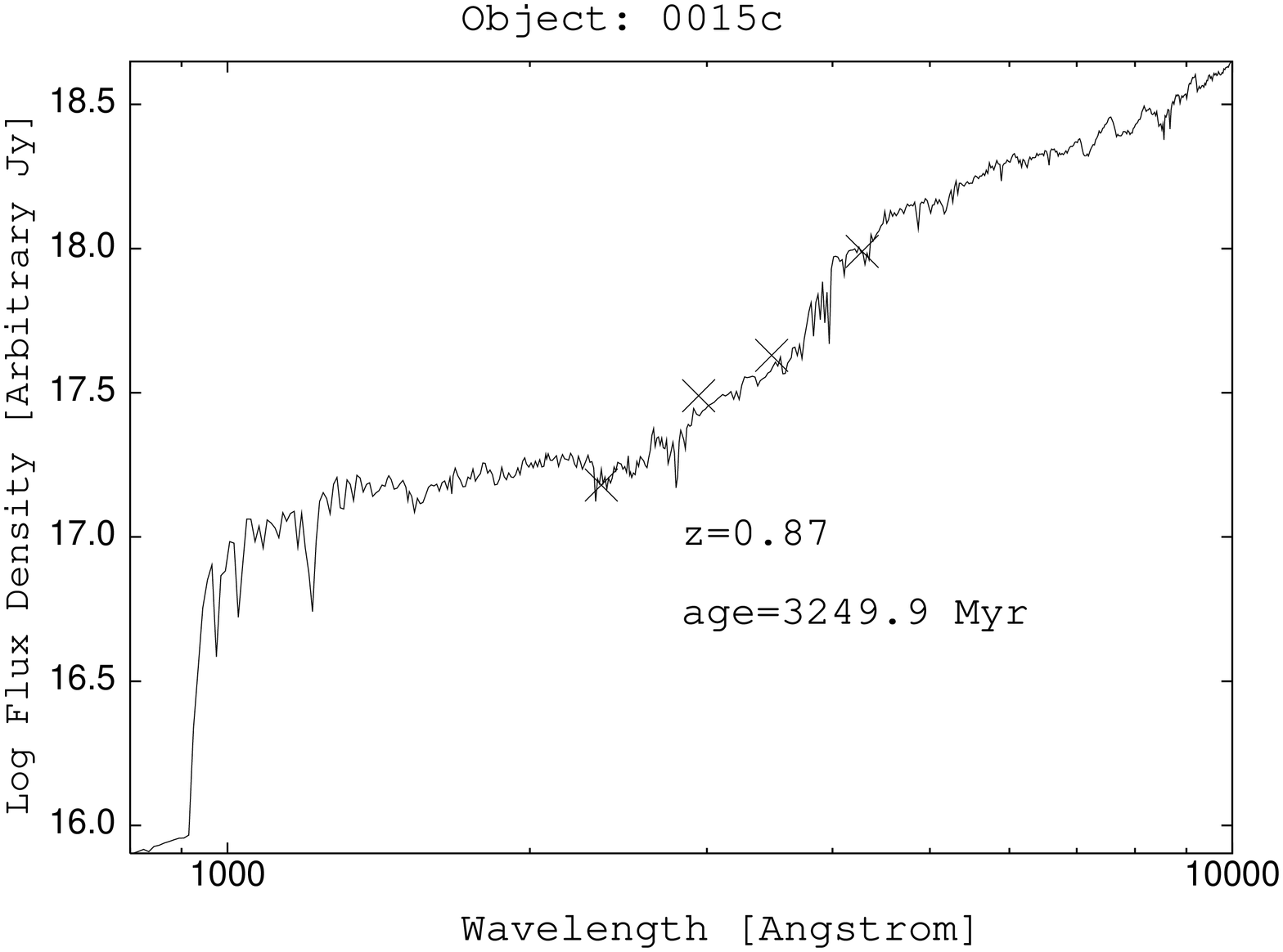,width=8cm,angle=-90}          
\mbox{\vspace*{0.5cm}}
\psfig{figure=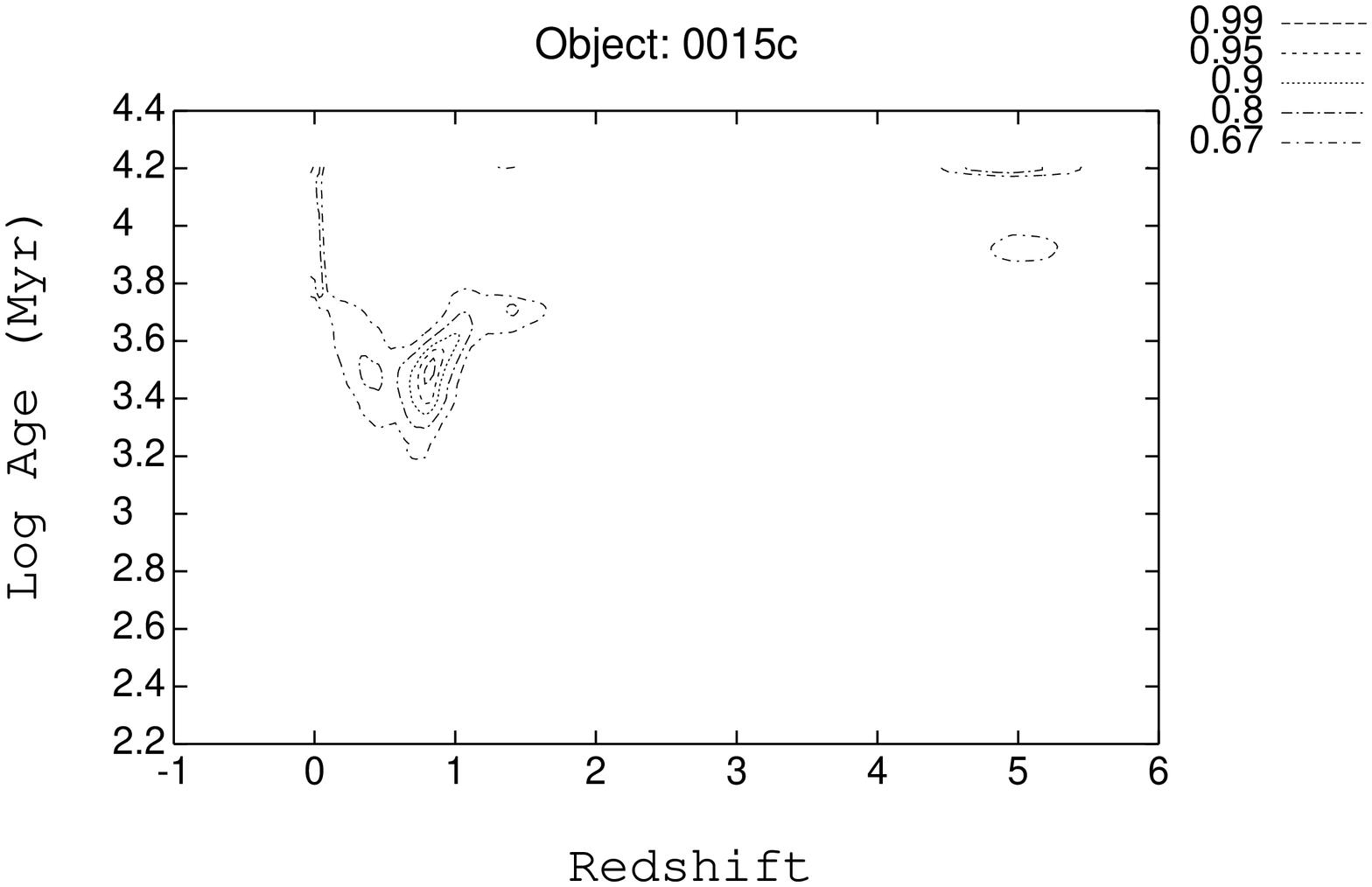,width=9cm,angle=-90,bbllx=520pt,bblly=90pt,bburx=68pt,bblly=745pt,clip=}
}}
\end{figure*}

\begin{figure*}[!h]
\centerline{PEGASE models for RC J0034+0513} 
\centerline{
\hbox{
\mbox{\hspace*{0.5cm}}
\psfig{figure=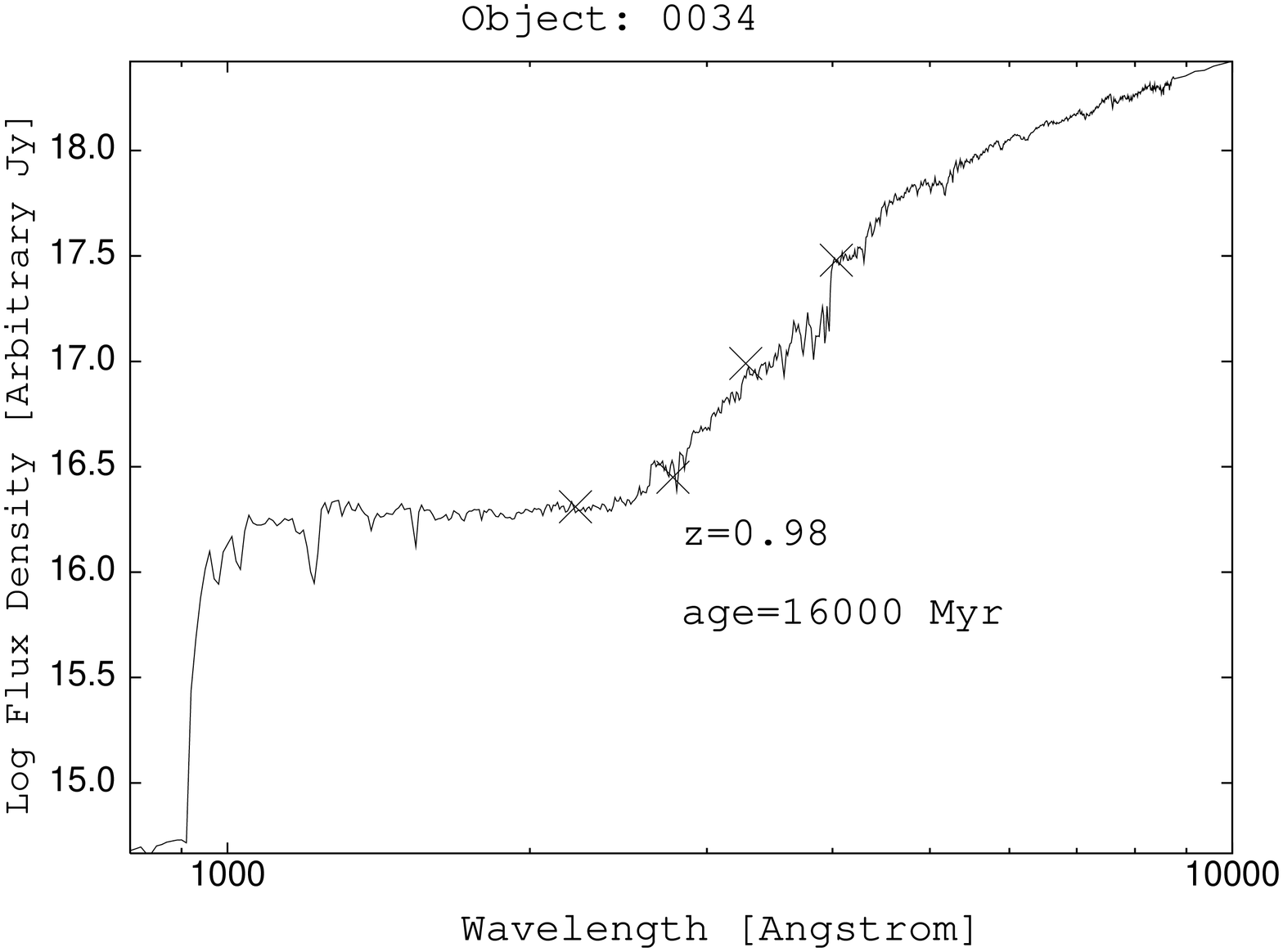,width=8cm,angle=-90}                   
\vspace*{0.5cm}
\psfig{figure=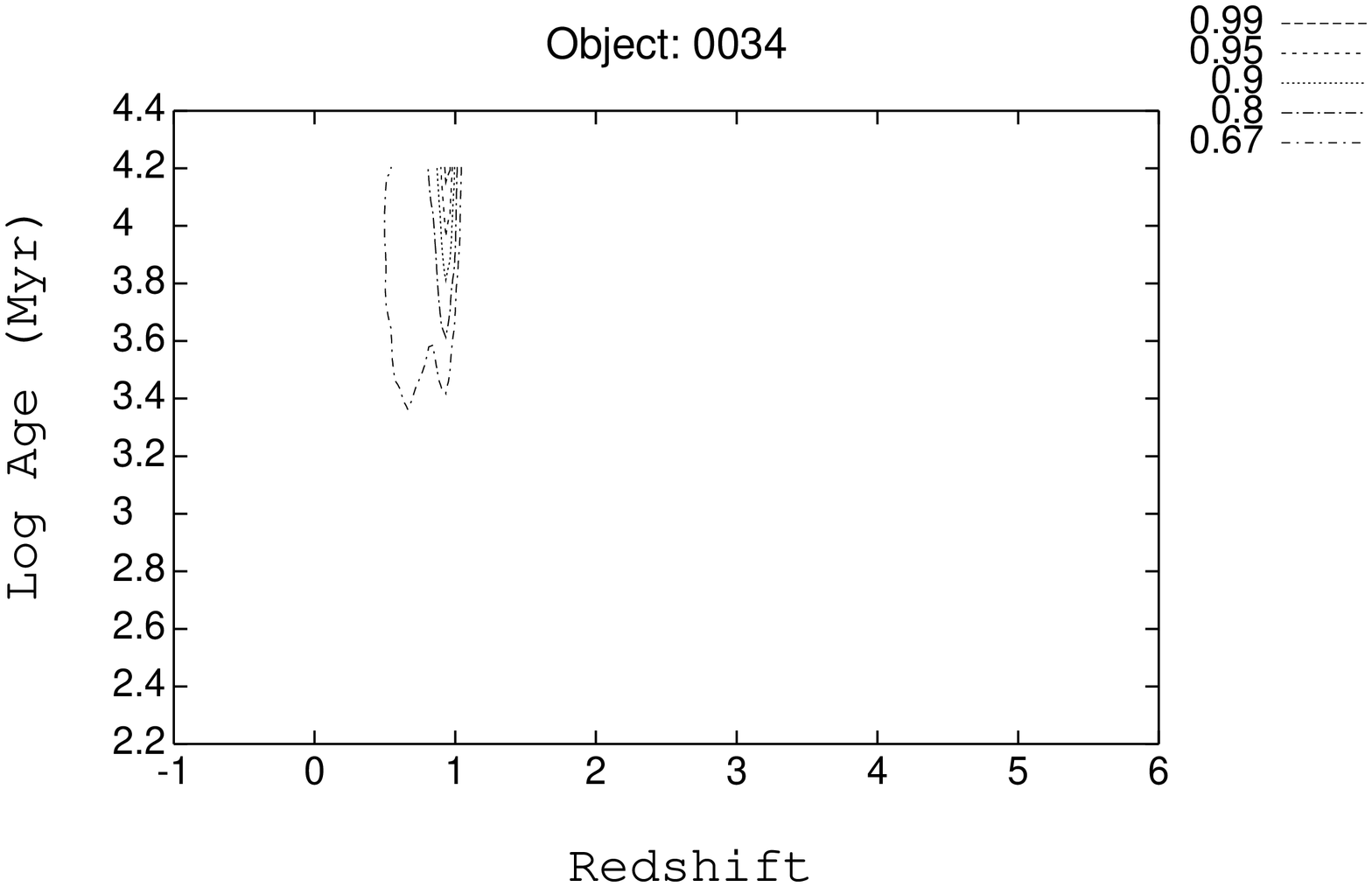,width=9cm,angle=-90,bbllx=520pt,bblly=90pt,bburx=68pt,bblly=745pt,clip=}
}}
\vspace*{4cm}
\end{figure*}

\end{document}